\documentclass[fleqn,usenatbib]{mnras}

\usepackage{newtxtext,newtxmath}

\usepackage[T1]{fontenc}

\DeclareRobustCommand{\VAN}[3]{#2}
\let\VANthebibliography\thebibliography
\def\thebibliography{\DeclareRobustCommand{\VAN}[3]{##3}\VANthebibliography}

\usepackage{graphicx}	
\usepackage{amsmath}	
\usepackage{xcolor}

\newcommand\msun{$\rm M_{\odot}$}

\title[The dark side of galaxy stellar populations]{The dark side of galaxy stellar populations II: 
The dependence of star formation histories on halo mass and on the scatter of the main sequence.}

\author[L. Scholz-D\'iaz et al.]{
Laura Scholz-D\'iaz,$^{1,2}$\thanks{E-mail: scholz@iac.es}
Ignacio Mart\'in-Navarro,$^{1,2}$
and Jes\'us Falc\'on-Barroso$^{1,2}$
\\
$^{1}$ Instituto de Astrofísica de Canarias, E-38205 La Laguna, Tenerife, Spain\\
$^{2}$ Universidad de La Laguna, Dept. Astrofísica, E-38206 La Laguna, Tenerife, Spain
}

\date{Accepted XXX. Received YYY; in original form ZZZ}

\pubyear{2021}

\begin{document}
\label{firstpage}
\pagerange{\pageref{firstpage}--\pageref{lastpage}}
\maketitle

\begin{abstract}
Nearby galaxies are the end result of their cosmological evolution, which is predicted to be influenced by the growth of their host dark matter halos. This co-evolution potentially leaves signatures in present-day observed galaxy properties, which might be essential to further understand how the growth and properties of galaxies are connected to those of their host halos. In this work, we study the evolutionary histories of nearby galaxies both in terms of their host halos and the scatter of the star-forming main sequence by investigating their time-resolved stellar populations using absorption optical spectra drawn from the Sloan Digital Sky Survey. We find that galaxy star formation histories depend on the masses of their host halos, and hence they shape the evolution of the star-forming main sequence over cosmic time. 
Additionally, we also find that the scatter around the z=0 star-forming main sequence is not (entirely) stochastic, as galaxies with currently different star formation rates have experienced, on average, different star formation histories. Our findings suggest that dark matter halos might play a key role in modulating the evolution of star formation in galaxies, and thus of the main sequence, and further demonstrate that galaxies at different evolutionary stages contribute to the observed scatter of this relation. 

\end{abstract}




\section{Introduction}
\label{sec:intro}

In the favored cosmological paradigm, the formation and growth of galaxies is predicted to be closely connected to the growth of their host dark matter halos \citep[][]{2018ARA&A..56..435W}. It is thought that halo mass growth is to first order the primary driver of galaxy formation \citep[e.g.,][]{white1978core,1984Natur.311..517B}, yet the baryonic cycle of galaxies also plays a key role in their evolution. As an example, in state-of-the art cosmological hydrodynamical simulations \citep[e.g.,][]{2014MNRAS.444.1453D, 2015MNRAS.446..521S, 2018MNRAS.473.4077P} the baryonic processes involved in galaxy formation are essential ingredients to be able to reproduce the populations, properties and scaling relations of observed galaxies. 

In this cosmological scenario, nearby galaxies are the end result of the interplay between their baryonic content and  their host halos over cosmic time, nonetheless, how the growth and properties of galaxies are connected to different halo properties is currently not fully understood \citep[][]{2018ARA&A..56..435W}. This work is part of a paper series which aims to probe the galaxy-halo connection by studying galaxy stellar populations in terms of their host dark matter halos. In \citet{2022MNRAS.tmp..362S} (Paper I) we investigated how average ages, [M/H] and [Mg/Fe] abundances of nearby galaxies depend on the masses of their host halos. At fixed halo mass, we recover well-known scaling relations of these parameters with stellar mass ($M_{\star}$) or velocity dispersion (at 1 effective radius, $\sigma_e$) \citep[e.g.,][]{1973ApJ...179..731F,1989PhDT.......149P,1992ApJ...398...69W,1993ApJ...411..153B,2000AJ....119.1645T,2005MNRAS.362...41G,2005ApJ...621..673T,2015MNRAS.448.3484M}, finding that galaxies become older and more metal-rich with increasing $M_{\star}$ or $\sigma_e$. But most interestingly, we also found a secondary dependence of the ages and [M/H] on the mass of their host halos (at fixed $M_{\star}$ or $\sigma_e$). To complement this previous work, in this follow-up paper we also explore the time evolution of the stellar populations of these nearby galaxies by studying their star formation histories (SFHs) and their potential dependence on halo mass. The predictive power of halo properties determining the evolutionary histories of galaxies has been studied recently by \citet[][]{2022arXiv220312702C} using neural networks and hydro-simulations, with their model being able to reproduce key aspects of the galaxy-halo connection.

Stellar populations studies are based on the fact that the cosmological evolution of galaxies leaves signatures in present day observed galaxy properties. In this sense, the stellar populations of galaxies in the Local Universe can be studied as fossil records of their past histories of star formation and chemical enrichment. 
This so-called `archaeological approach' \citep[e.g.,][]{2005ApJ...621..673T} allows to trace the evolution of individual galaxies by recovering their SFHs, i.e., how the star formation rates (SFRs) of the galaxies evolve with cosmic time. The reconstruction of galaxy SFHs can be done through spectroscopy either by using specific spectral features \citep[e.g.,][]{2003MNRAS.341...33K,2005MNRAS.362...41G} or the full spectral range \citep[e.g.,][]{2004Natur.428..625H,2005MNRAS.358..363C,2006MNRAS.365...46O, 2007MNRAS.381.1252T,2015MNRAS.448.3484M,2016MNRAS.463.2799I}, and either by assuming a particular shape of the SFH \citep[e.g.,][]{2003MNRAS.341...33K,2005MNRAS.362...41G,2007MNRAS.381.1252T} or   in a non-parametric manner \citep[e.g.,][]{2007MNRAS.378.1550P,2005MNRAS.358..363C,2006MNRAS.365...46O,2015MNRAS.448.3484M,2016MNRAS.463.2799I,2018Natur.553..307M,2019MNRAS.487.4939M}. Full-spectral fitting schemes usually find the optimal linear combination of single stellar population models (SSPs) that best fit the observed spectrum of galaxy. This allows to determine the stellar mass associated to stellar populations of different ages, and hence, to obtain a non-parametric galaxy SFH. Additionally, galaxy SFHs obtained through these fossil records have provided important constrains on their evolution. It has been found that the SFHs of the galaxies depend on their stellar mass \citep[e.g.,][]{2004Natur.428..625H, 2007MNRAS.378.1550P, 2015MNRAS.448.3484M,2016MNRAS.463.2799I,2018ApJ...861...13C}. More massive galaxies build the bulk of their stellar mass rapidly at early times, having a high SFR peak at that time, while less massive galaxies have more extended SFHs and form over a more prolonged period of time. This so-called `downsizing' of massive galaxies is also reflected in observed [$\rm \alpha$/Fe] abundance ratio, with more massive galaxies being also more $\rm \alpha$-enhanced \citep[e.g.,][]{1999MNRAS.306..607J,2000AJ....119.1645T,2000AJ....120..165T,2005ApJ...621..673T,2006MNRAS.370.1106G,2010MNRAS.408...97K,2015MNRAS.448.3484M}. This $\rm \alpha$-enhancement in massive galaxies indicates that their star formation have been truncated earlier than for less massive galaxies, with the latter having more extended star formation histories \citep{2005ApJ...621..673T,2011MNRAS.418L..74D}.

On the other hand, the evolution of galaxy SFRs over cosmic time has been traditionally studied through the `Star-forming Main Sequence' (SFMS) \citep[e.g.,][]{2004MNRAS.351.1151B,2007ApJ...660L..43N,2007ApJS..173..267S,2012ApJ...754L..29W,2015ApJ...801L..29R}. The SFMS is a tight scaling relation (0.2-0.3 dex) for star-forming galaxies that consists of a linear correlation between integrated star formation rates (SFRs) and stellar masses of the galaxies (in logarithmic scale). These galaxy SFRs are typically derived through spectroscopy, but in this case using emission lines (e.g. the $\rm H\alpha$ line), which are sensitive to the presence of short-lived massive young stars in the galaxies \citep[e.g.,][]{2013seg..book..419C}. Moreover, a SFMS is also found for star-forming galaxies at higher redshifts, and this relation evolves over cosmic time \citep[e.g.,][]{2014ApJS..214...15S,2017MNRAS.470..651R}, with its normalization decreasing with time. In addition, the SFMS and its evolution with cosmic time have also been studied through the fossil record approach \citep[][]{2019MNRAS.482.1557S}.


While star-forming galaxies follow the SFMS, there is also a population of galaxies with lower SFRs that fall below this relation and are classified as quiescent galaxies. 
The fraction of quiescent galaxies increases with cosmic time, dominating the high mass end of the stellar mass function at z=0 \citep[e.g.,][]{2013A&A...556A..55I,2013ApJ...777...18M}. Therefore, it is often suggested that star-forming galaxies evolve into quiescent ones \citep[e.g.,][]{2004ApJ...608..752B,2007ApJ...665..265F}, assuming that galaxies at different redshifts are representative of the same galaxy population seen at different cosmic times, i.e., higher-z galaxies are the progenitors of galaxies in the Local Universe. This transition in which galaxies cease their star formation is referred to as `quenching'.

The existence of these massive quiescent galaxies in the Local Universe, which typically have old stellar populations and formed at early epochs and on short timescales, seems to challenge a naive interpretation of the hierarchical scenario of the $\rm\Lambda$CDM cosmology. In this picture, the merging of small galaxies with continuous gas accretion and recycling over cosmic time would result in the formation of massive young galaxies in the local Universe, which are not observed today \citep[e.g.,][]{1984ApJ...287..586B,1996A&A...311..425B,1999ApJ...525..144V,2015MNRAS.448.3484M,2018MNRAS.475.3700M,2020A&A...644A.117L}. The quenching mechanisms able to shut down star formation in these galaxies early on and keep them quiescent over cosmic time are still under debate. 

In this work, we employ the fossil records approach to derive SFHs and SFRs of central galaxies in the Local Universe through their absorption spectra. We study the connection between galaxies and their host halos by exploring the potential dependence of galaxy SFHs on halo mass. Additionally, we investigating how the scatter of the SFMS might be modulated by the different evolutionary histories of the galaxies. 

The paper is organized as follows: In section \ref{sec:data} we introduce the SDSS data, the group catalog and the galaxy samples used in this work. Section \ref{analysis} describes the kinematic and stellar population analyses, and how we determine galaxy SFRs and SFHs. The assembly time of the galaxies across the stellar-to-halo mass relation and the velocity dispersion - halo mass relation are shown in  section \ref{res:t50}. The dependence of the cumulative mass distributions and star formation histories of the galaxies on their host dark matter halos and on the scatter of the SFMS are shown in sections \ref{res:sfh_mh} and \ref{res:SFMS_sfh}, respectively. In section \ref{sec:quench} we show the the evolutionary tracks of galaxies across the $\rm SFR-M_{\star}$ relation. Our results are discussed in section \ref{sec:discussion} and summarized in section \ref{sec:concl}.

\section{Sample and data}
\label{sec:data}

Here, we give a brief overview of the sample and data used in our analysis. For a more detailed description, we refer the reader to  \citet{2022MNRAS.tmp..362S}, which presents our main galaxy sample and the data used in this paper series.

Our main galaxy sample consists of 8,801 central galaxies from the group and cluster catalog from \citet{2007ApJ...671..153Y} with $0.01\!<\!z\!<\!0.2$. In this catalog the dark matter halo masses of the groups/clusters are estimated through abundance matching by ranking the total luminosity of the groups, and we assigned these halo masses to their corresponding central galaxies. We selected only centrals in groups/clusters with more than three members to have more reliable halo masses. We also connected these halo estimates with optical/NIR spectra (3800-9200 \AA) from the Sloan Digital Sky Survey Data Release 9 \citep[SDSS DR9;][]{2000AJ....120.1579Y,2012ApJS..203...21A}. We derived galaxy stellar population properties from these spectra: the velocity dispersion at 1 $R_e$, stellar masses, mass-weighted average ages, metallicities [M/H], and [Mg/Fe] abundances. Our central galaxies have stellar masses between $10^{9.7}$ and $10^{11.8}$ \msun, and halo masses between $10^{11.7}$ and $10^{15.4}$ \msun.

In this work we measure star formation rates through the star formation histories of the galaxies (see section \ref{an_SFR} for more details). We compare those measurements with the ones from \citet{2004MNRAS.351.1151B}. These authors estimate galaxy SFRs using nebular emission lines, and for galaxies with weak emission or with active galactic nucleii (AGN) the SFRs are estimated using the 4000-\AA$\,$break (see \citet{2004MNRAS.351.1151B} for more detailed description). Given that we do not use halo masses in this comparison, we use an extended sample in which we do not apply any cuts in the number of galaxies in the group/cluster. This extended sample consists of 271,450 galaxies of which a subsample of 153,489 galaxies are in common \citet{2004MNRAS.351.1151B}.

\section{Analysis}
\label{analysis}
We performed kinematic and stellar population analyses using full-spectral fitting, as briefly described below in section \ref{an_kinepop}. We refer the reader to  \citet{2022MNRAS.tmp..362S} for a more detailed description of the method. This approach allowed us to determine average stellar population properties, star formation histories and present-day star formation rates, as described in   \citet{2022MNRAS.tmp..362S}, and sections \ref{an_sfh} and \ref{an_SFR}.

\subsection{Kinematic and stellar populations analyses}
\label{an_kinepop}
We derived galaxy kinematics and stellar population properties using the penalized pixel-fitting algorithm pPXF \citep[pPXF, ][]{2004PASP..116..138C,2017MNRAS.466..798C} fed with the $\alpha$-variable MILES stellar population synthesis models \citep{2010MNRAS.404.1639V,2015MNRAS.449.1177V}.  pPXF finds the linear combination of single stellar population (SSP) models that best reproduce a galaxy spectrum by convolving the SSP models with a line-of-sight velocity distribution (LOSVD) parametrized by a Gauss-Hermite expansion \citep{1993ApJ...407..525V}. We employed models of 15 different ages (from 0.03 to 13.5 Gyr) sampled roughly logarithmically at young ages, and with a 1 Gyr step between them at old ages. We selected models with 11 different metallicities [M/H] (-2.27 to +0.4 dex) and the two available [$\alpha$/Fe] abundances (+0.0 and +0.04 dex). These models adopt an universal Kroupa IMF \citep{2001MNRAS.322..231K} and $\alpha$-variable isochrones \citep{2004ApJ...612..168P,2006ApJ...642..797P}. In the fitting process, we first measured the stellar and gas kinematics, and then kept the stellar kinematics fixed in the stellar population analysis to avoid the velocity dispersion-metallicity degeneracy \citep{2011MNRAS.415..709S}. We note that we fitted the spectra from 4800 to 5400 \AA, where the magnesium triplet 5167.7 \AA$\,$dominates the sensibility to $\alpha$-elements. Therefore, hereafter we will employ the notation [Mg/Fe] instead of [$\alpha$/Fe]. We also imposed linear regularization \citep{10.5555/1403886} to the pPXF solution of the stellar population analysis in order to find the smoothest one among the many degenerate solutions compatible with the observed spectrum \citep[see][]{2017MNRAS.466..798C}. 

\subsection{Star formation histories}
\label{an_sfh}

\begin{figure}
    \centering
    \includegraphics[scale = 0.44]{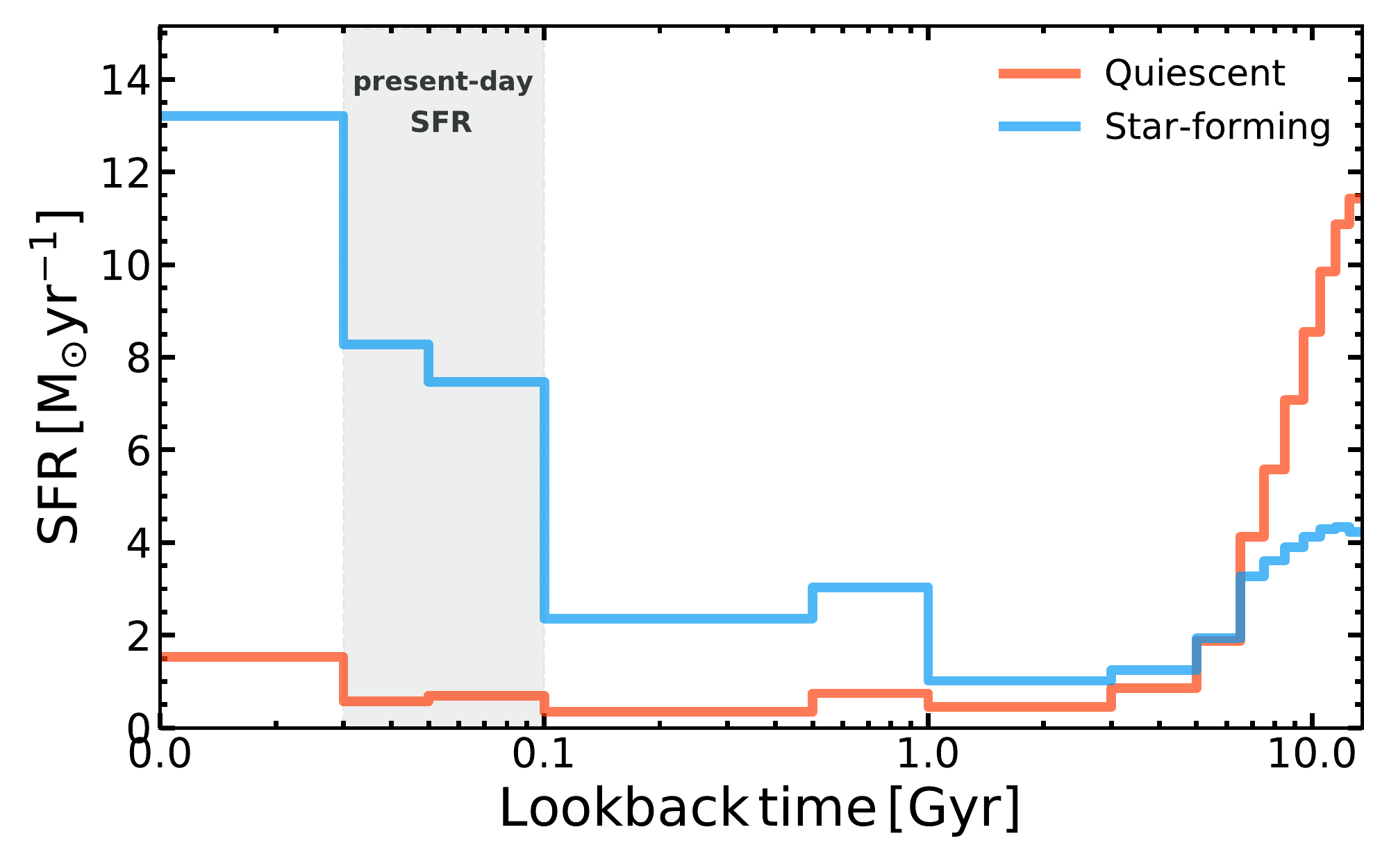}
    \caption{Different galaxy star formation histories. Star formation rate vs. the age of the stellar populations for a galaxy that has assembled early and fast (orange line), one that formed late and over a long period of time (blue line). The grey bandwidth indicates we use to define the galaxy SFRs (see section \ref{an_SFR}).} 
    \label{fig:sfh_cartoon}
\end{figure}

 In the stellar population analysis, pPXF is fed with SSP models that are sampled in a grid of ages, [M/H], and [$\alpha$/Fe] abundances, so that the best-fitting solution corresponds to a series weights assigned to each SSP. The SSP models are scaled to one solar mass, and hence, the weights associated to SSPs of different ages can be translated into stellar mass fractions using mass-to-light ratios. In another words, we can derive the masses associated to the stellar populations of different ages present in the galaxies (i.e., the mass formed in the galaxies at different times). Thus, combining these masses with the time-steps between SSPs allows us to estimate the SFRs of the galaxies at different times, i.e, their SFHs. Therefore, these SFHs are sampled discretely by construction, as their time resolution is given by the ages of the SSPs.  
 
 Note that with this method we are not able to distinguish between the stellar mass formed in-situ and ex-situ (e.g., stellar mass originated from galaxy mergers) and assumes that the total stellar mass of a galaxy has formed in-situ \citep[][]{2020MNRAS.491..823B}. We discuss this caveat in more detail in section \ref{sec:disc_cav}

Fig. \ref{fig:sfh_cartoon} exemplifies two different galaxy SFHs. We show the star formation rate as a function of the age of the stellar populations, where different SFHs are shown in different colors. Focusing on the SFH shown in orange, we show the SFH of a galaxy that has assembled fast and early. We observe that this galaxy had high SFRs in the past over a relatively short period of time and then had almost ceased its star formation activity until present-day. On the other hand, in blue we show the SFH of a galaxy that has assembled late and over a long period of time. This galaxy had also a peak of SFR at early times, although with a lower level of SFR (compared to the one shown in orange). Then, it had a minimum of SFR activity, followed by a steadily increase of SFR over cosmic time. Note that in the following sections we show the SFHs lookback time (LBT) in linear scale, but we have employed a logarithmic scale for the LBT in these examples in order to better visualize the region in which we calculate the SFRs (see section \ref{an_SFR}. Moreover, although in these examples we plotted the SFHs using step functions in order to show the time-steps of the models, in the rest of the figures we used solid lines in order to better visualize and disentangle the different SFHs shown in the panels.

\subsection{Star formation rates}
\label{an_SFR}

\begin{figure}
    \centering
    \includegraphics[scale = 0.435]{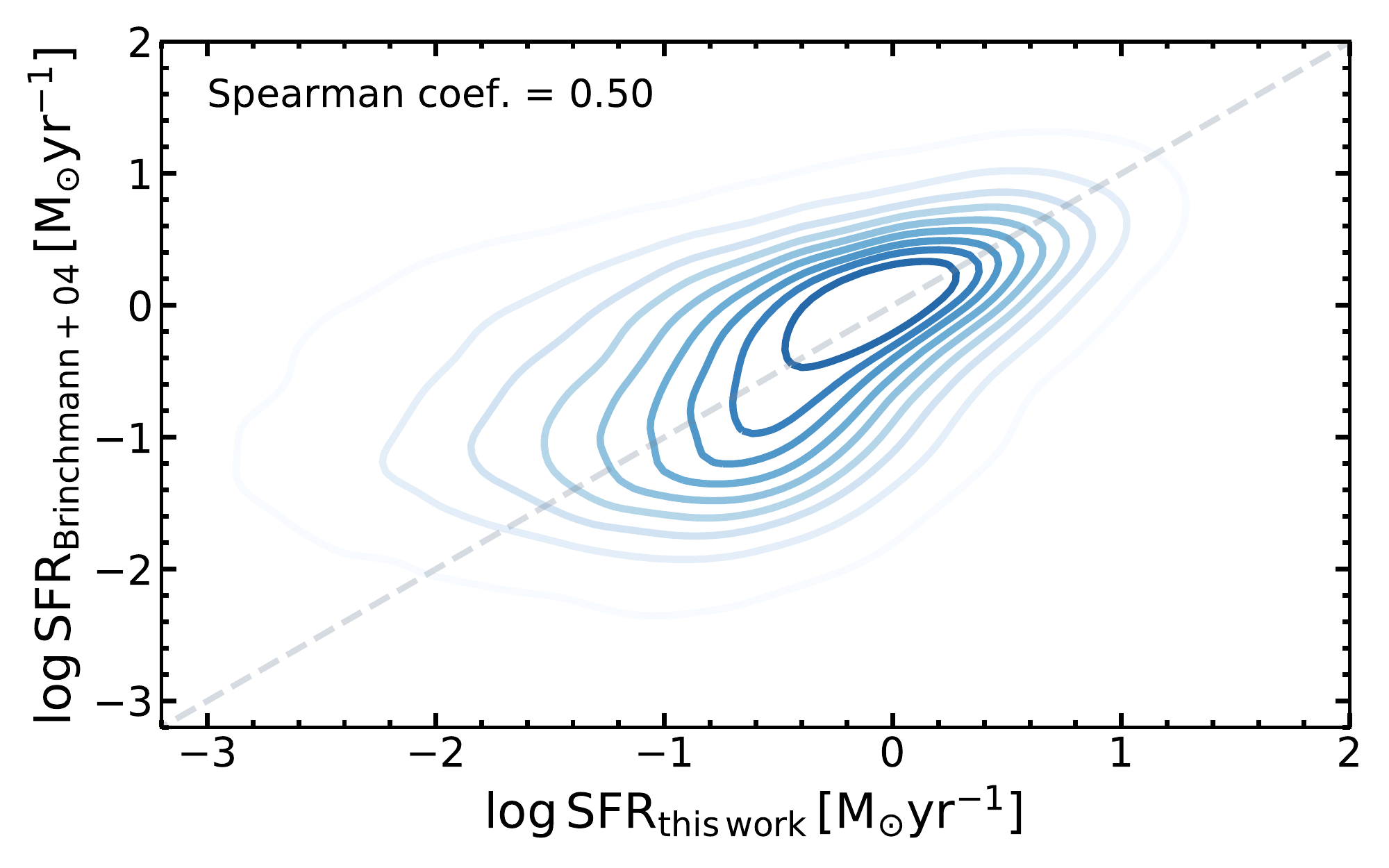}
    \caption{Comparison between different star formation rate measurements. The colored contours show the SFRs measured using either nebular emission lines or the D4000\AA$\,$ if the galaxies have AGN or weak emission \citep{2004MNRAS.351.1151B} vs. SFRs obtained through SFHs measured from absorption spectra (this work). Our SFRs have been shifted -0.5 dex in log-space to account for the offset between the two measurements. The grey dashed line shows the 1:1 relation for reference. } 
    \label{fig:comp_brinch}
\end{figure}

In order to study the connection between the scatter of the SFMS and galaxy SFHs, we need measurements of present-day SFRs in order to classify the galaxies according to their position respect to the SFMS. Thus, we calculate the present-day SFR for the best-fitting SFHs by extracting the stellar mass formed over the last 70 Myr, which corresponds to the last two ages in our SSP grid (excluding the youngest models). This time-step was defined so that the our SFRs are consistent with those of \citet{2004MNRAS.351.1151B}, which are estimated independently using either nebular emission lines or the D4000\AA$\,$ if the galaxies have AGN or weak emission. Figure \ref{fig:sfh_cartoon} shows the time-step where our SFRs are measured as a grey bandwidth over different galaxy SFHs. Note that we exclude from our definition the last SFR point, which corresponds to the mass associated to the model of 30 Myr (the youngest SSPs), given that very young SSPs are less reliable. By doing so, we trace better the SFRs from \citet{2004MNRAS.351.1151B}, although with an offset of 0.5 dex between these two SFR estimators, with our measurements being shifted towards higher SFRs.

To define our present-day SFR, we used our extended galaxy sample (which also includes galaxies in isolation and in groups with 3 or less members) to have larger statistics and a more complete sample, as we do not need halo masses for this analysis. In Fig. \ref{fig:comp_brinch} we show the comparison between our 
SFRs and the ones from \citet{2004MNRAS.351.1151B} for the subsample of galaxies in common. The colored contours show the SFRs from \citet{2004MNRAS.351.1151B} vs. our SFRs measured from absorption spectra (which have been shifted -0.5 dex in log-space to account for the offset between the two measurements). The grey dashed line corresponds to the 1:1 relation for reference. Note that galaxies without recent star formation do not appear in this figure by construction, given that these galaxies have zero stellar mass formed recently $(\rm SFR_{70 \, Myr} = 0)$. We find that our SFRs correlate with the ones from \citet{2004MNRAS.351.1151B} with a Spearman correlation coefficient of 0.5, although we note the 0.5 dex offset towards higher SFRs. Even so, we would like to highlight this good agreement, given that these SFRs are obtained with completely different approaches, which have their own systematic errors, caveats and limitations, and are also sensitive to different star-formation time-scales. Our method is not sensitive to very low levels of SFR, which effectively translate into  zero SFR, while the method based on emission lines is sensitive to lower and more instantaneous SFRs \citep[see section 1.2.1 of the review from][]{2013seg..book..419C}.

\subsection{Star-forming main sequence}
\label{sec:an_SFMS}

\begin{figure}
    \centering
    \includegraphics[scale = 0.385]{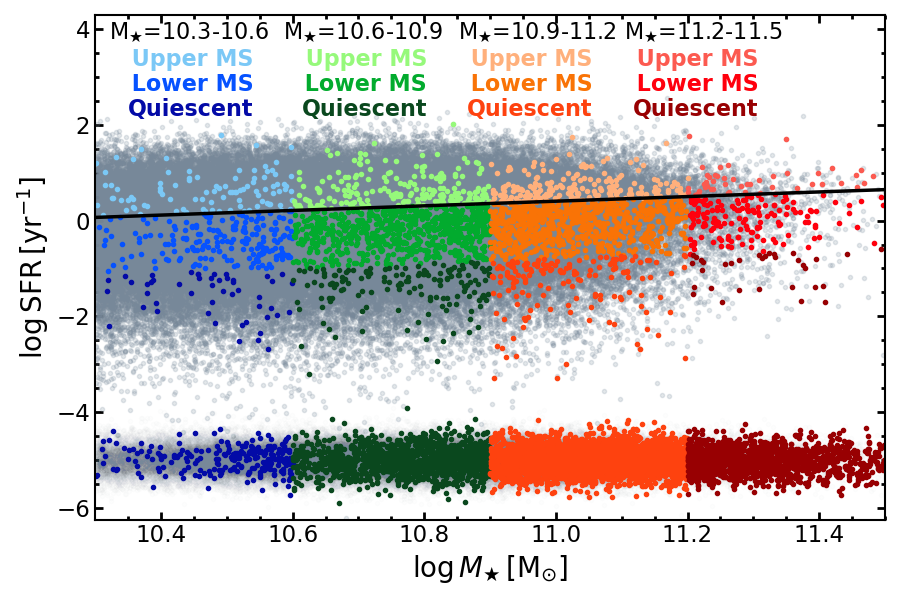}
    \caption{$\rm  SFR$-$ \rm  M_{\star}$ relation for all galaxy types, where circles indicate individual galaxies, and the SFMS resulting from the analysis described in section \ref{sec:an_SFMS} is shown as a black solid line. Grey circles correspond to galaxies from the extended sample and colored ones to main sample galaxies. Galaxies in different stellar mass bins are shown in different colors and within each bin, galaxies in the upper MS are shown in a light color, the ones in the lower MS with an intermediate color and quiescent ones with a dark color (for more details see section \ref{sec:an_SFMS}). For visualization purposes, quiescent galaxies with $\rm SFR_{70 \, Myr}=0$ are artificially plotted with arbitrarily low SFR values of $\rm \log \, SFR=-5 \ dex$ randomly scattered with a standard deviation of 0.25 dex. }
    \label{fig:sfms}
\end{figure}

\begin{figure}
    \centering
    \includegraphics[scale = 0.405]{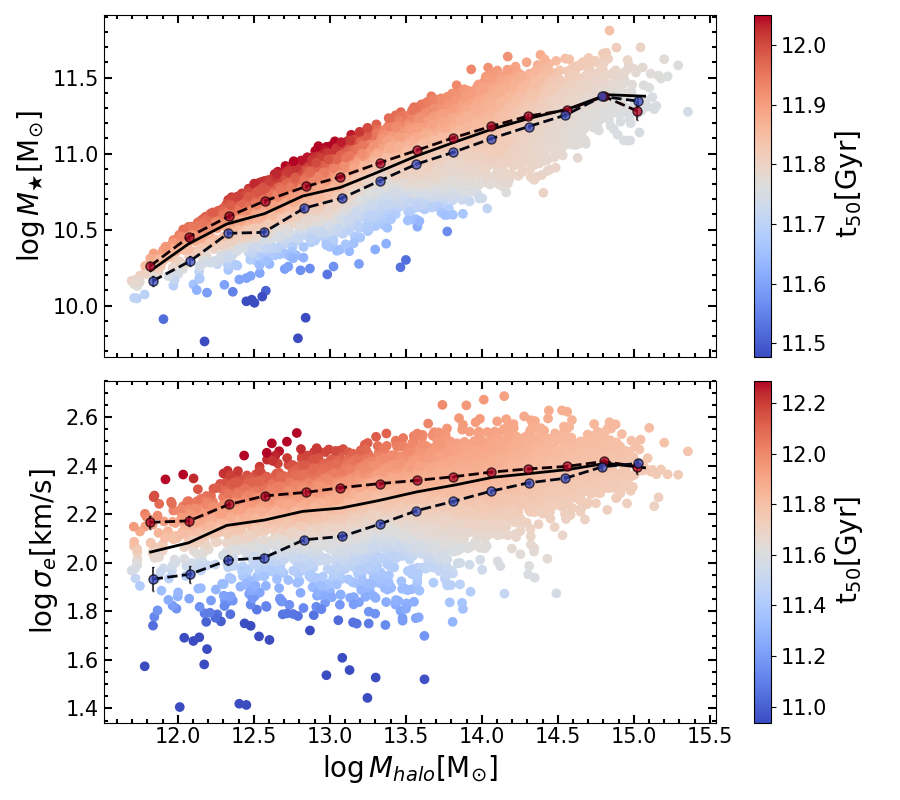}
    \caption{Stellar-to-halo mass relation (upper panel) and velocity dispersion (measured within 1 $R_e$) as a function of halo mass (bottom panel) for SDSS central galaxies.  Both relations are color-coded by the lookback time in which the galaxies have formed half of their present stellar mass, $t_{50}$, resulting from our stellar population analysis described in section \ref{analysis}. We applied the LOESS algorithm to the color-coding in order to highlight the global trends across the corresponding relation. The median relation are shown as a black solid lines for reference. At fixed $M_h$, red and blue circles connected by dashed lines indicate the mean $M_{\star}$ (upper panel) or $\sigma_e$ (bottom panel) of galaxies above the $\rm75^{th}$-percentile and below $\rm25^{th}$-percentile of the $t_{50}$ distribution. The scatter of the SHMR correlates with the time in which the galaxies form half of their present-day stellar mass at the low halo mass regime.} 
    \label{fig:shmr_vdhmr_t50}
\end{figure}

\begin{figure*}
    \centering
    \includegraphics[scale = 0.59]{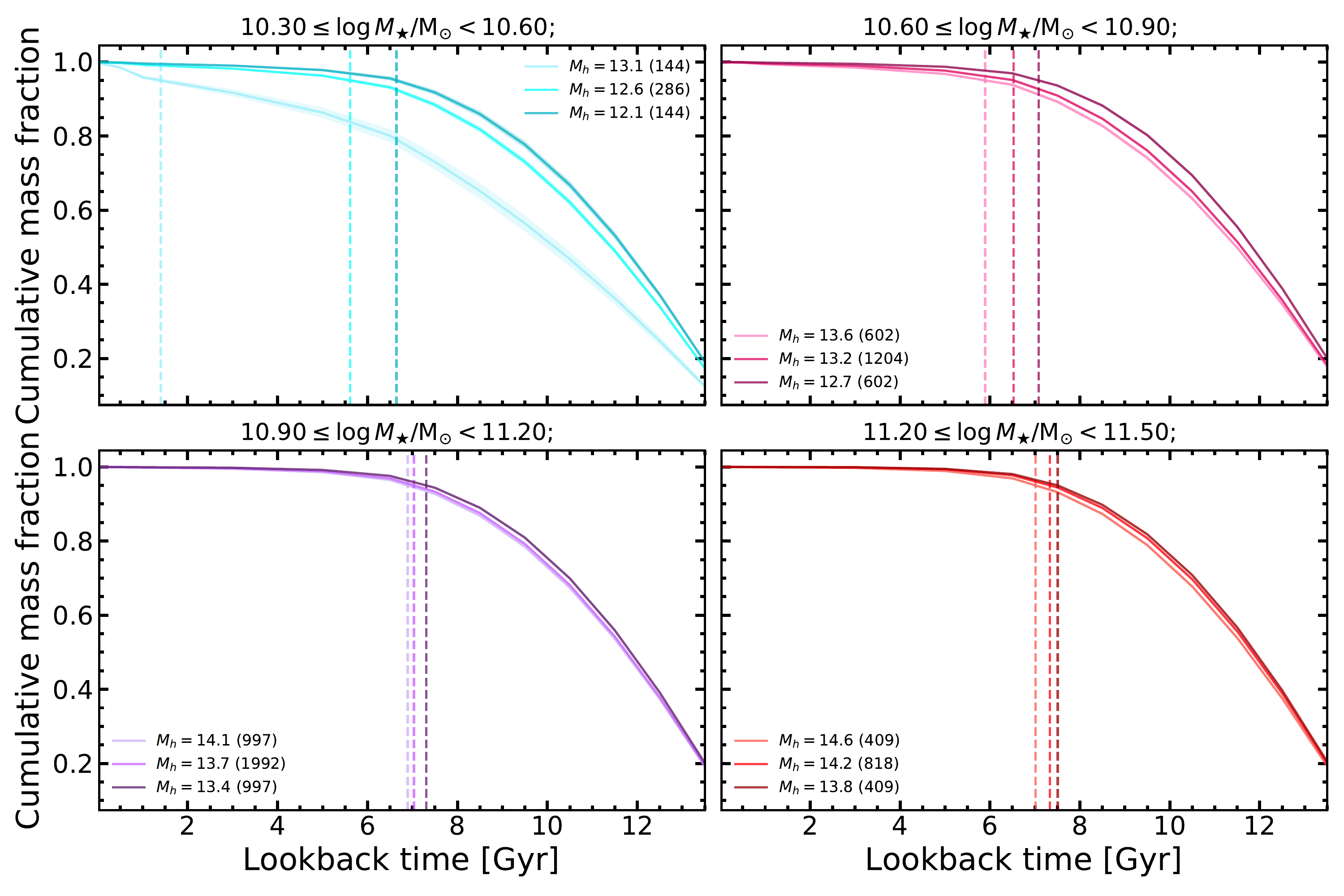}
    \caption{The dependence on halo mass of the cumulative stellar mass fraction in different stellar mass bins: $10.3\!<\!\log M_{\star}/ \rm M_{\odot} \!\leq\! 10.6$ (upper left panel), $10.6\!<\!\log  M_{\star}/ \rm M_{\odot} \!\leq\! 10.9$ (upper right panel), $10.9\!<\!\log  M_{\star} / \rm M_{\odot} \!\leq\! 11.2$ (bottom left panel) and $11.2\!<\!\log  M_{\star} /\rm M_{\odot} \!\leq\! 11.5$ (bottom right panel). In each panel, we show the time evolution of the mean cumulative mass fraction for different halo mass bins indicated as solid lines, where lighter colors correspond to larger halo masses: $\rm 25^{th}$~pctl. $<M_h$ (dark),  $ \rm 25^{th}$~pctl. $\leq M_h<$ $\rm 75^{th}$~pctl. (medium) and  $M_h \leq$ $\rm 75^{th}$~pctl (light). Colored regions indicate the 1$\sigma$ uncertainty of the mean. For reference, vertical dashed lines indicate when 95\% of the stellar mass is reached. Low stellar mass galaxies form the bulk of their stars earlier and faster if they reside in less massive halos (at fixed $M_{\star}$) than if they live more massive halos. }
    \label{fig:cum_mass_mh}
\end{figure*}

In this section we describe how we derive the SFMS for star-forming galaxies, as we use this relation as a reference to study the connection between galaxy SFHs and the scatter across the SFR-$\rm M_{\star}$ relation. 

Similarly to section \ref{an_SFR}, in this analysis we also use our extended galaxy sample to have larger statistics. Moreover, we selected only galaxies with $M_{\star}  \! \geq  \! 10^{10.3} \, \rm M_{\odot}$ and  $M_{\star} \! \leq  \! 10^{11.5} \, \rm M_{\odot}$ for consistency with the analyses shown in the following sections. In order to derive the SFMS, we fit the relation that star-forming galaxies follow in the $\rm SFR$-$ \rm  M_{\star}$ relation. Thus, we first select only  star-forming galaxies as the ones with a specific star formation rate (sSFR) above $\rm 10^{-11.1} \, M_{\odot} yr^{-1}$ . We justify this threshold choice in appendix \ref{ap:sel_sf}. Note that this selection of star-forming galaxies is only employed to compute the SFMS and is not further used in the paper. Then, once the star-forming galaxies are selected, we fit linearly the $\rm SFR$-$ \rm  M_{\star}$ relation for these galaxies only, and we define our SFMS as the best-fitting solution:

\begin{equation}
     \rm \log SFR = 0.48^{+0.49}_{+0.47} \times \log M_{\star} - 4.92^{-4.77}_{-5.03}
\end{equation}

In Fig. \ref{fig:sfms} we show the $\rm  SFR$-$ \rm  M_{\star}$ relation for all galaxy types, where circles indicate individual galaxies and the SFMS is shown as a black solid line.  Grey circles correspond to galaxies from the extended sample, while colored circles correspond to galaxies from our main sample (which have more reliable halo mass estimations). 

In section \ref{res:SFMS_sfh} we study the relation between the scatter of the SFMS and the SFHs of the galaxies. For that, we classify the galaxies according according to their position with respect to the main sequence (MS), at fixed stellar mass. In Fig. \ref{fig:sfms} different colors indicate different stellar mass bins. Within these bins we classify as upper MS galaxies the ones that are above the SFMS, as lower MS the ones that are below the SFMS with $\rm |SFMS-logSFR|> 1.2 \, dex$, and as quiescent the ones below the SFMS with $\rm |SFMS-logSFR| < 1.2 \, dex$ or that have $\rm SFR_{70 \, Myr}$ = 0. Galaxies in these categories are distinguished in Fig. \ref{fig:sfms} with different color shades. Lighter colors correspond to upper MS galaxies, intermediate colors to lower MS galaxies and and darker colors to quiescent galaxies (within each stellar mass bin). Quiescent galaxies with $\rm SFR_{70 \, Myr}$ = 0 would not appear in this plot by construction, therefore for visualization purposes we artificially plot them with arbitrarily low SFR values of $\rm \log \, SFR=-5 \ dex$ randomly scattered with a standard deviation of 0.25 dex. Note that we do not use these artificial values for any computation performed throughout this work.

We also tested the robustness of our aproach and our SFR measurements by repeating the same analysis, but in this case using  SFRs from \citet{2004MNRAS.351.1151B} to compute the SFMS and classify the galaxies. We also repeated this test using stellar masses from \citet[][]{2004MNRAS.351.1151B}. In both cases we find consistent results and in agreement with the ones presented in section \ref{res:SFMS_sfh}.

\section{Galaxy assembly time across the SHMR and the VDHMR}
\label{res:t50}

\begin{figure*}
    \centering
    \includegraphics[scale = 0.59]{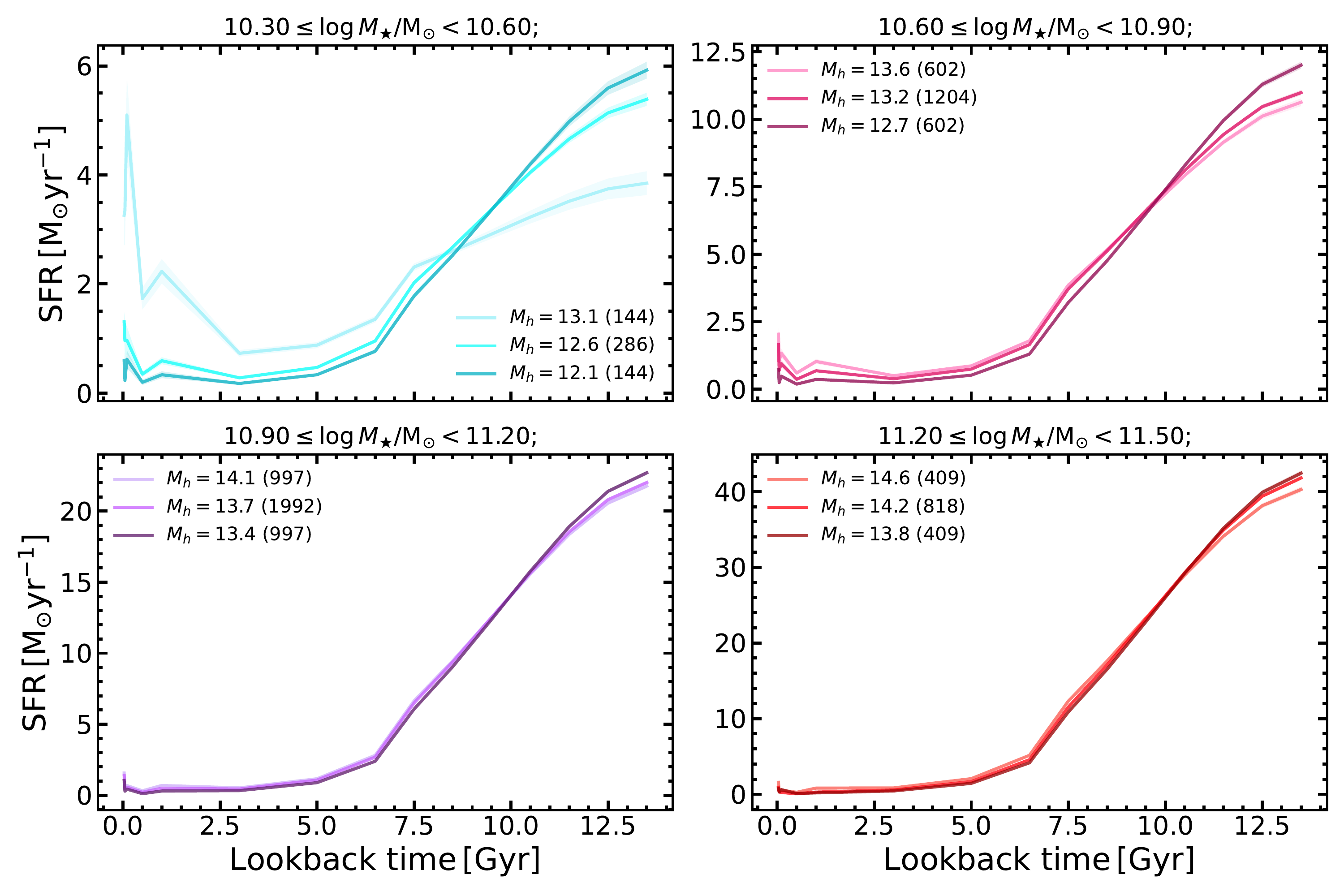}
    \caption{The dependence on halo mass of the star formation histories of the galaxies in different stellar mass bins: $10.3\!<\!\log M_{\star}/ \rm M_{\odot} \!\leq\! 10.6$ (upper left panel), $10.6\!<\!\log  M_{\star}/ \rm M_{\odot} \!\leq\! 10.9$ (upper right panel), $10.9\!<\!\log  M_{\star} / \rm M_{\odot} \!\leq\! 11.2$ (bottom left panel) and $11.2\!<\!\log  M_{\star} /\rm M_{\odot} \!\leq\! 11.5$ (bottom right panel). In each panel, we show the time evolution of the mean SFR of the galaxies for different halo mass bins indicated as solid lines, where lighter colors correspond to larger halo masses: $\rm 25^{th}$~pctl. $<M_h$ (dark),  $ \rm 25^{th}$~pctl. $\leq M_h<$ $\rm 75^{th}$~pctl. (medium) and  $M_h \leq$ $\rm 75^{th}$~pctl (light). Colored regions indicate the 1$\sigma$ uncertainty of the mean. Low stellar mass galaxies hosted by less massive halos (at fixed $M_{\star}$) had higher SFRs at earlier times, and end up with low SFRs today. In contrast, the ones residing in more massive halos had lower SFRs in the past, and higher levels of SFR today.}
    \label{fig:sfh_mh}
\end{figure*}

In Paper I we studied how average stellar properties behave across the stellar-to-halo mass relation (SHMR) and the velocity dispersion - halo mass relation (VDHMR) for central galaxies of our main galaxy sample. In this paper we complement this previous work by studying the time-resolved stellar populations of these galaxies, being able to derive the evolution of their stellar mass until present-day, as described in section \ref{an_sfh}. In this section we follow a similar approach to the one shown in Paper I by investigating the behavior of the assembly time of the galaxies across the SHMR and VDHMR. For that, we compute the lookback time  in which these galaxies form half of their present stellar mass, $t_{50}$, with galaxies of larger $t_{50}$ forming earlier than galaxies with lower $t_{50}$. Note that we do not aim to set absolute constraints on the $t_{50}$ values, but the goal of this analysis is to compare the galaxies in a relative way, as we are interested in the relative difference of the SFHs of different galaxies. 

In Fig. \ref{fig:shmr_vdhmr_t50} we show the SHMR (upper panel) and the VDHMR (bottom panel) color-coded by the looback time in which the galaxies formed half of their present stellar mass ($t_{50}$), which results from our stellar population analysis described in section \ref{analysis}. To highlight the global trends, we applied the locally weighted regression (LOESS) algorithm of \citet{doi:10.1080/01621459.1988.10478639} to the color-coding of Fig.~\ref{fig:shmr_vdhmr_t50}\footnote{We use the LOESS implementation by  \citet{2013MNRAS.432.1862C} }. The median relations are shown as a black solid lines for reference. At fixed $M_h$, red and blue circles connected by dashed lines indicate the mean $M_{\star}$ (upper panel) or $\sigma_e$ (bottom panel) of galaxies above the $\rm75^{th}$-percentile and below $\rm25^{th}$-percentile of the $t_{50}$ distribution.

We first note that the trends of $t_{50}$ across both relations resemble the ones of average age (Figs. 1 and 2 of Paper I), which is expected, given that the average age of a galaxy corresponds to the average of the stellar populations of different ages present in the galaxy, weighted by their mass. In this regard, galaxies that are older on average also form the bulk of the stars at earlier times compared to younger galaxies, which assemble over longer time-scales. Hence, as seen for the age, the scatter of both relations correlate with $t_{50}$ in the low halo mass regime, while the trend seems to disappear at higher halo masses. At low halo masses, and at fixed halo mass, galaxies assemble their stellar mass earlier as their stellar mass increases. We also looked into the SFHs of the galaxies (at fixed halo mass), checking that we recover well-known trends of the SFHs with stellar mass and the downsizing of massive galaxies (see section \ref{sec:intro}), as we observe that more massive galaxies form earlier and faster, having higher SFR at earlier times than less massive galaxies, which have more extended SFHs. 

At the same time, we also find that at fixed stellar mass, galaxies form the bulk of their stars earlier if their are hosted by lower mass halos, specially for low stellar masses.

\section{Dependence of the SFHs on halo mass}
\label{res:sfh_mh}

\begin{figure*}
    \centering
    \includegraphics[scale = 0.59]{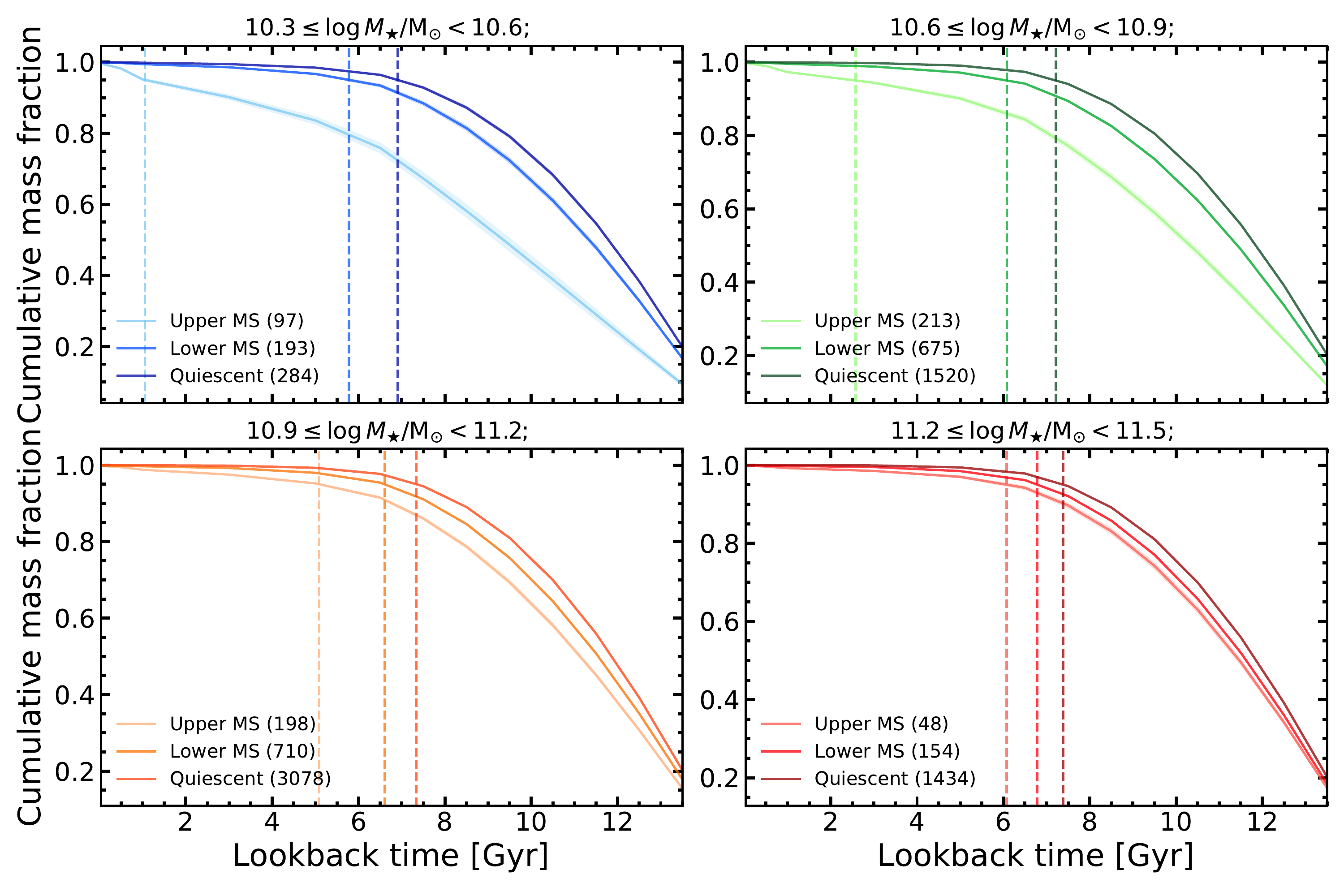}
    \caption{The dependence of the cumulative stellar mass fraction on the scatter of the SFMS in different stellar mass bins: $10.3\!<\!\log M_{\star}/ \rm M_{\odot} \!\leq\! 10.6$ (upper left panel), $10.6\!<\!\log  M_{\star}/ \rm M_{\odot} \!\leq\! 10.9$ (upper right panel), $10.9\!<\!\log  M_{\star} / \rm M_{\odot} \!\leq\! 11.2$ (bottom left panel) and $11.2\!<\!\log  M_{\star} /\rm M_{\odot} \!\leq\! 11.5$ (bottom right panel). In each panel, we show the time evolution of the mean cumulative mass fraction (solid lines) for galaxies in the upper MS (light color), in the lower MS (medium color) and quiescent galaxies (dark color). Colored regions indicate the 1$\sigma$ uncertainty of the mean. For reference, vertical dashed lines indicate when 95\% of the stellar mass is reached. Galaxies below the MS (at fixed $M_{\star}$) assemble  on short time-scales and earlier compared to galaxies above the MS, specially at the low stellar mass regime. }
    \label{fig:cum_mass_sfms}
\end{figure*}

\begin{figure*}
    \centering
    \includegraphics[scale = 0.59]{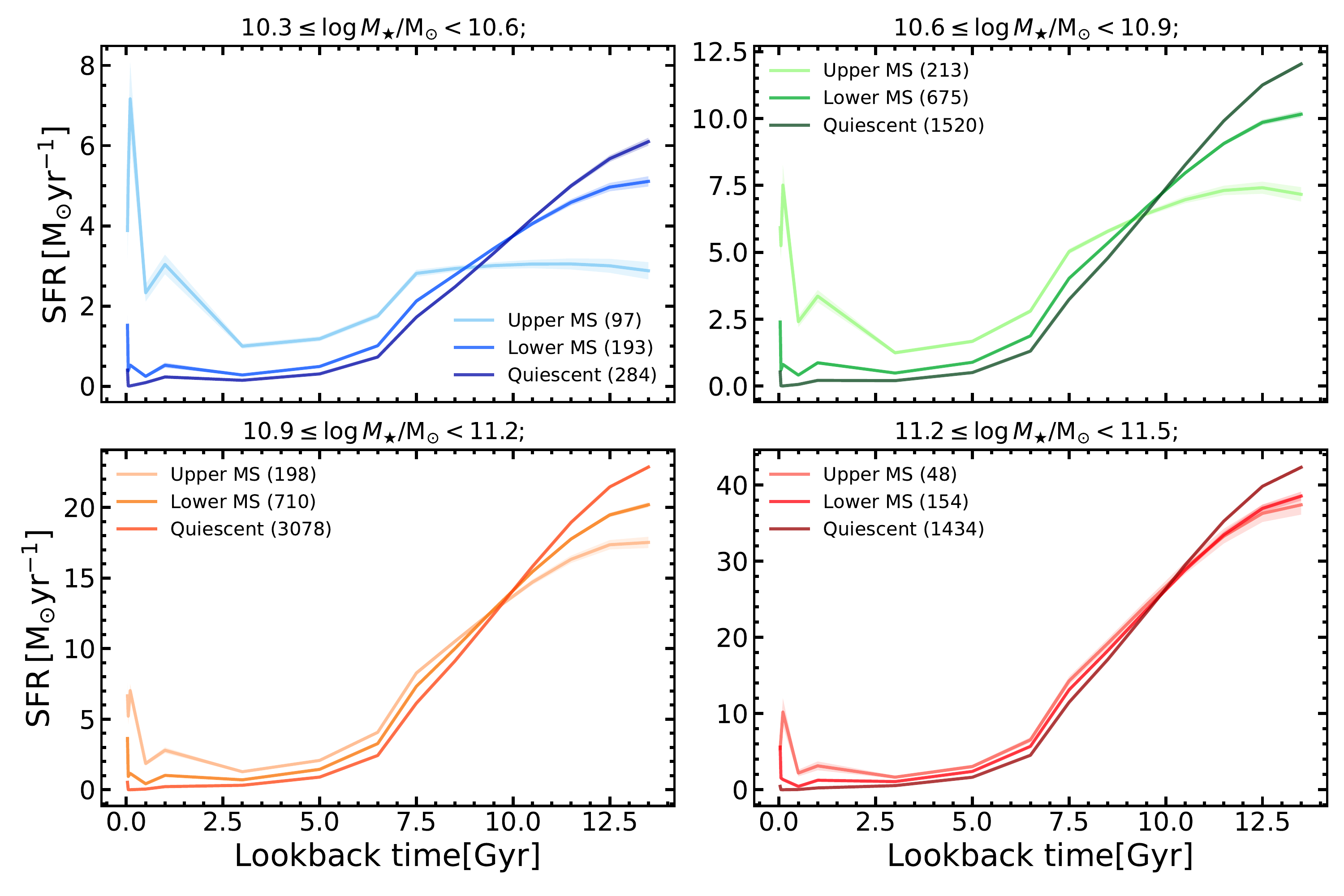}
    \caption{The dependence of galaxy star formation histories on the scatter of the SFMS in different stellar mass bins: $10.3\!<\!\log M_{\star}/ \rm M_{\odot} \!\leq\! 10.6$ (upper left panel), $10.6\!<\!\log  M_{\star}/ \rm M_{\odot} \!\leq\! 10.9$ (upper right panel), $10.9\!<\!\log  M_{\star} / \rm M_{\odot} \!\leq\! 11.2$ (bottom left panel) and $11.2\!<\!\log  M_{\star} /\rm M_{\odot} \!\leq\! 11.5$ (bottom right panel). In each panel, we show the time evolution of the mean SFR of the galaxies (solid lines) for galaxies in the upper MS (light color), in the lowe MS (medium color) and quiescent galaxies (dark color). Colored regions indicate the 1$\sigma$ uncertainty of the mean. Galaxies with low SFRs today (at fixed $M_{\star}$) had higher SFRs at earlier times, while the ones with high SFRr had lower levels of SFR in the past.}
    \label{fig:sfh_sfms}
\end{figure*}

In section \ref{res:t50} we found that the assembly time of the galaxies seem to be modulated by the halo mass of their host dark matter halos. In this section we further investigate the dependence of galaxy evolutionary histories on halo mass. 

In order to isolate the effect of halo mass on the galaxy SFHs, we first need to account for the dependence on stellar mass. For that, we first divide our main galaxy sample into four stellar mass bins: $ \log M_{\star}/ \rm M_{\odot} \in$  (10.3, 10.6], (10.6, 10.9], (10.9, 11.2] and (11.2, 11.5]. We have excluded the lowest and most massive galaxies, as we have significantly less galaxies in these two regimes. Then, each bin is subdivided into three halo mass bins using the $ \rm 25^{th}$- and $\rm 75^{th}$ percentiles of the $M_h$ distribution: $\rm 25^{th}$~pctl. $<M_h$,  $ \rm 25^{th}$~pctl. $\leq M_h<$ $\rm 75^{th}$~pctl. and  $M_h \leq$ $\rm 75^{th}$~pctl. When we compute average quantities for each $M_h$ bin (e.g., the average SFH), we follow \citet[][]{2013MNRAS.432..359T} by assigning a weight to each galaxy in the bin to account for the dependence on stellar mass. The weights are given by the ratio between a target gaussian distribution centered in the mean $M_{\star}$ of the stellar mass bin, and a fit to the original stellar mass distribution of the $M_h$ bin (for each galaxy, the ratio is evaluated in the stellar mass of the galaxy). In this sense, we are able to compare equivalent bins in terms of stellar mass without reducing the number of galaxies.

Figure \ref{fig:cum_mass_mh} shows the evolution of the cumulative stellar mass fraction of the galaxies over cosmic time, where different panels correspond to the different stellar mass bins. In each panel, solid lines indicate the time evolution of the mean cumulative mass fraction for galaxies in the different halo mass bins, where lighter colors indicate larger mean halo masses. Colored regions indicate the 1$\sigma$ uncertainty of the mean. For reference, vertical dashed lines indicate when 95\% of the stellar mass is reached. Focusing first on the lowest stellar mass bin (upper left), we observe that galaxies in less massive halos (dark cyan) assemble earlier and faster than galaxies in more massive halos (light cyan). Galaxies in halos of intermediate masses (medium cyan) correspond to an intermediate case between the other two. In this sense, the 95\% of the stellar mass is reached later than for galaxies in less massive halos, which is shown with vertical dashed lines in the panels. As we increase the stellar mass, we see on the other panels the same general behavior, although the relative differences between the different halo mass bins are reduced. For the most massive stellar mass bins (bottom panels) the differences in $t_{95}$ are significantly smaller. Hence, the dependence of the cumulative mass growth on halo mass seems to be more significant for galaxies with low stellar masses. For the lowest stellar mass bin, galaxies in more massive halos assemble the 95\% of their mass at a look back time ($t_{95}$) of 1.4 Gyr (light cyan), galaxies in intermediate halos at 5.6 Gyr (intermediate cyan), while galaxies in less massive halos at 6.6 Gyr (dark cyan). Similarly, galaxies in the stellar mass bin of (10.6, 10.9] have $t_{95}$ of 5.9 (light pink), 6.5 (intermediate pink), and 7.1 Gyr (dark pink). In the case of the stellar mass bin of (10.9, 11.2], $t_{95}$ correspond to 6.7 (light purple), 7 (intermediate purple), and 7.3 Gyr (dark purple). While for the highest stellar mass bin, galaxies have more similar $t_{95}$ of 7 (light red), 7.3 (intermediate red) and 7.5 (dark red).

Figure \ref{fig:sfh_mh} is analogous to Fig. \ref{fig:cum_mass_mh}, but solid lines indicate the time evolution of the mean SFR of the galaxies in different halo mass bins. We observe again the largest differences between the halo mass bins for the lowest $M_{\star}$ bin (upper left panel). We find that galaxies in less massive halos (dark cyan) have a the largest SFR peak at early times, followed by a rapid decrease of their SFR activity with cosmic time. For younger ages (LBT < 4 Gyr) galaxies hosted by less massive halos exhibit, on average, low SFR values compared to galaxies living in more massive halos. This is in agreement with Fig. \ref{fig:cum_mass_mh}, as these galaxies have formed the bulk of their stellar mass earlier and faster. In contrast, galaxies in more massive halos (light blue) had lower SFR at early times, then have gradually decreased their SFR. Then, around 2.5 Gyr ago, the SFR has generally increased and has the highest values at present day. In agreement with Fig. \ref{fig:cum_mass_mh}, these galaxies have formed the bulk of their stellar mass more gradually and over a longer period of time. As before, the intermediate halo mass bin  (medium blue) corresponds to an intermediate case between the previous two. The peak of SFR at early times it is slightly lower than the one shown in dark cyan, then the SFR decrease not so rapidly and it keeps higher levels of SFR at present day. Similarly to Fig. \ref{fig:cum_mass_mh}, the same general behavior is also seen for the other stellar mass bins, although the relative differences between the different halo mass bins are reduced with increasing $M_{\star}$. We also see that the most massive stellar mass bins show the smaller relative differences. \\

We found that halo mass influences galaxy star formation histories. This effect is more noticeable for low mass galaxies, although we cannot conclude from our analysis whether the baryonic cycle in low mass galaxies is more heavily influenced by the mass of dark matter halos or our results for higher stellar masses just reflect the limitations of our stellar population modelling (see section \ref{sec:disc_cav}). Given the correlation between stellar mass and halo mass (SHMR), note that galaxies in the lowest stellar mass bin are hosted by halos at the low halo mass regime. Moreover, in Paper I we analyzed average  stellar population properties across the stellar-to-halo mass relation. For low halo masses ($M_h< \rm10^{13.5}$) and at fixed stellar mass, we found that galaxies in halos with different masses have also different ages and [M/H], with galaxies in less massive halos being older and more metal-rich than galaxies in more massive ones. This is in agreement with the findings presented in Fig. \ref{fig:cum_mass_mh} and \ref{fig:sfh_mh}, as galaxies assembled earlier and faster in less massive halos at fixed stellar mass. Hence, these galaxies end up with older populations and are chemically more evolved at present day.

\section{Dependence of the SFHs on the scatter across the $\rm SFR-M_{\star}$ relation}
\label{res:SFMS_sfh}

\begin{figure*}
    \centering
    \includegraphics[scale = 0.6]{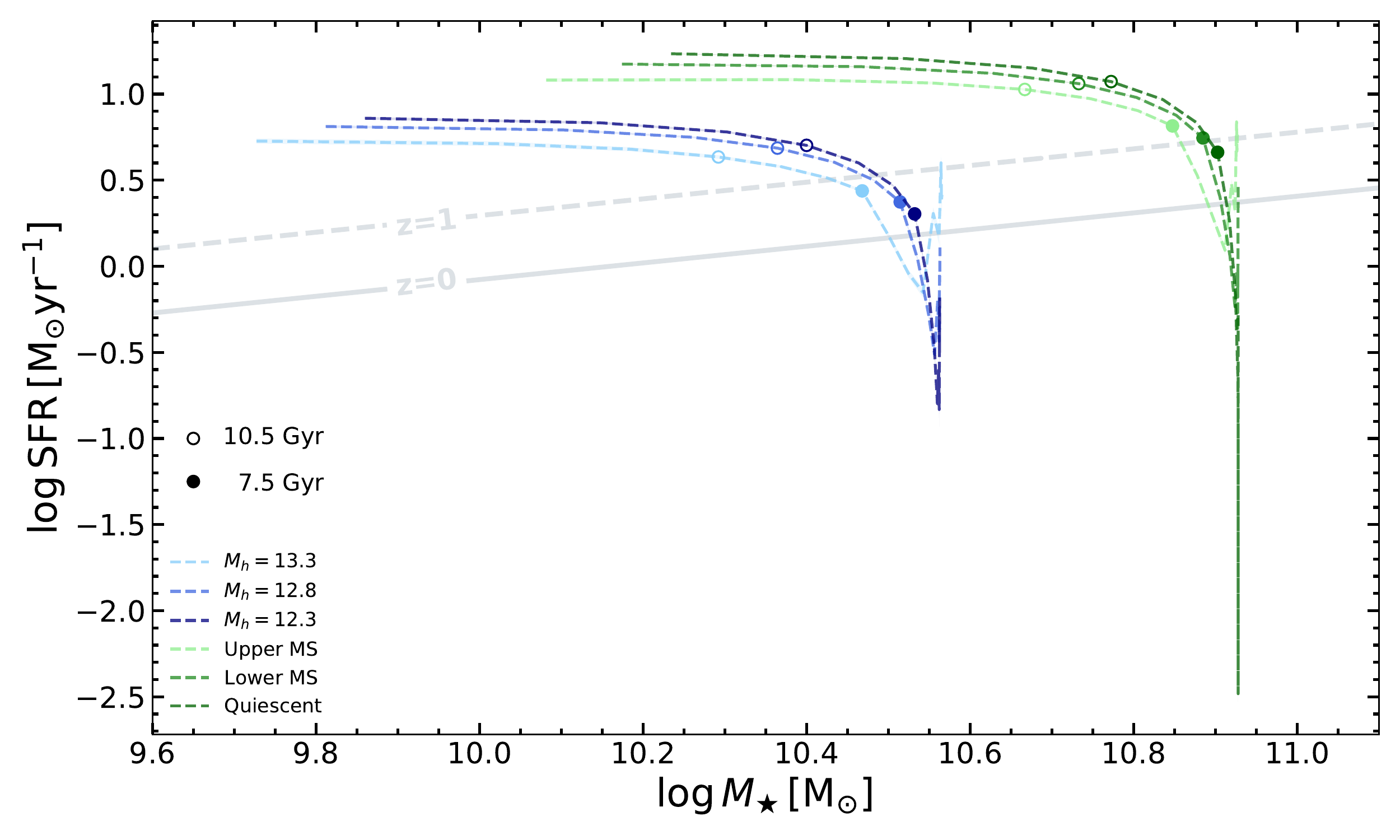}
    \caption{Evolution of the galaxies across the $\rm  SFR$-$ \rm  M_{\star}$ plane. The average tracks of the galaxies are indicated with dashed lines and different colors correspond to the low (blue) and high mass bins (green). The dashed lines of different color shades indicate different halo mass bins (within the low stellar mass bin) or bins of galaxies with different positions relative to the SFMS (within the high stellar mass bin). Filled circles indicate the position of the galaxies along the tracks at a lookback time of 10.5 (empty circles) and 7.5 Gyr (filled circles). The solid grey line correspond to the SFMS a $\rm z=0$ resulting from the analysis described in \ref{sec:an_SFMS}. The evolutionary tracks of the galaxies show a sudden quencing phase in which the SFR decrease abruptly.}
    \label{fig:quenching}
\end{figure*}

In this section we explore the connection between the scatter of the $\rm SFR-M_{\star}$ relation at present day and galaxy evolutionary histories. In particular, we investigate whether galaxies with different positions with respect to the SFMS at $\rm z=0$ have also different cumulative mass distributions and star formation histories.

First, we classify the galaxies according to their position with respect to the SFMS, differentiating between galaxies in the upper MS, lower MS and quiescent galaxies (see section \ref{sec:an_SFMS} for more details on the classification). We also subdivide the galaxies into the same stellar mass bins presented in section \ref{res:sfh_mh}. In this sense, there are three bins related to the position of the galaxies with respect to the SFMS within each stellar mass bin. To have equivalent bins in terms of stellar mass, we also weight their corresponding distributions as described in section \ref{res:sfh_mh}. In Fig. \ref{fig:sfms} we show the $\rm SFR$-$ \rm  M_{\star}$ relation where galaxies in the different bins shown with different colors (see section~\ref{sec:an_SFMS}).

In Fig.~\ref{fig:cum_mass_sfms} we investigate how the cumulative mass fraction of galaxies for the different stellar mass bins (different panels) is connected to the scatter of the $\rm SFR$-$ \rm  M_{\star}$  relation. Analogously to Fig. \ref{fig:cum_mass_mh}, each panel shows the time evolution of the mean cumulative mass fraction (solid lines) for galaxies in the upper MS (light color), in lower MS (medium color) and quiescent galaxies (dark color). We observe that the galaxy cumulative mass fractions depend on the position of a galaxy with respect to the SFMS for all stellar mass bins, although the differences are larger for lower stellar masses. Quiescent galaxies assemble earlier and faster than galaxies in the upper and lower MS, while galaxies in the upper MS form later and over a longer period of time. 

We also show the star formation histories of the galaxies in Fig.~\ref{fig:sfh_sfms}, which is analogous to Fig. \ref{fig:cum_mass_sfms}, but solid lines indicate the time evolution of the mean SFR of the galaxies for the upper MS, lower MS and quiescent galaxies. Similarly to what we found in Fig. \ref{fig:cum_mass_sfms}, we observe that the star formation histories of galaxies in different positions in the $\rm  SFR$-$ \rm  M_{\star}$ relation are also different. We see again that the difference is larger for the lowest stellar mass bins, but significant in all of them. Quiescent galaxies had the highest SFR at early epochs, which is followed by a rapid decrease of SFR. These galaxies keep low levels of SFR today. In contrast, galaxies in the upper MS had a lower peak of SFR at early times and the following decrease of SFR happens more gradually. Then, the SFR increase again at later times. Lower MS galaxies correspond to an intermediate case between the previous two. 

These results imply that galaxies at different positions relative to the $\rm  SFR$-$ \rm  M_{\star}$ have experienced considerably different SFHs, or in another words, that galaxies in similar regions of the $\rm  SFR$-$ \rm  M_{\star}$ relation also have similar SFHs.

\section{Evolution across the $\rm  SFR - M_{\star}$ relation}
\label{sec:quench}

In section \ref{res:sfh_mh} we found that the cumulative  mass growth and the SFHs of the galaxies depend on the mass of their host halos, at fixed $M_{\star}$. In addition, we found in section \ref{res:SFMS_sfh} that the scatter of the SFMS has an evolutionary component, as the present-day SFRs of the galaxies also depend on their SFHs and cumulative mass growth, for a given stellar mass. In this section, we take these analyses a step further. Similarly to \citet[][]{2019MNRAS.482.1557S}, the fossil record approach also allow us to measure the evolutionary tracks of individual galaxies across the $\rm  SFR - M_{\star}$, given that for each galaxy we are able to determine mass fractions and SFRs associated to populations of different ages, as described in section \ref{an_sfh}. 

We focus only on representative stellar mass regimes for each analysis in order to better visualize the tracks across the $\rm  SFR - M_{\star}$, as it is difficult to disentangle tracks of galaxies with similar present-day $M_{\star}$, and we have already assessed the dependence of the SFHs both on halo mass and on the scatter of the SFMS in previous sections. Therefore, we select two stellar mass bins, $\rm 10.3 \leq \log M_{\star} < 10.7$ and $10.7 \leq \log M_{\star} < 11.1$, and study the halo mass dependence in0 the lower stellar mass bin and dependence on scatter of the SFMS in the higher stellar mass bin. Thus, for the low mass bin, we subdivide the galaxies into bins of galaxies in halos of different masses, analogously to section \ref{res:sfh_mh}. Similarly, for the high mass bin we subdivide the galaxies into bins of galaxies with different positions relative to the SFMS  resulting from the analysis described in \ref{sec:an_SFMS}, analogously to section \ref{res:SFMS_sfh}. We chose galaxies in these stellar mass regimes, as  we show in sections \ref{res:sfh_mh} and \ref{res:SFMS_sfh} that the differences in the SFHs are more significant for low and intermediate masses, while they are less significant for more massive galaxies. For these two cases, we also follow the analysis described in section \ref{res:sfh_mh} to compare equivalent bins in terms of stellar mass.

Figure \ref{fig:quenching} shows the average tracks across the $\rm  SFR - M_{\star}$ relation over cosmic time for galaxies in the different bins. The tracks are indicated with dashed lines and different colors correspond to the low (blue) and high mass bins (green). The dashed lines of different color shades indicate different halo mass bins (within the low stellar mass bin) or bins of galaxies with different positions relative to the SFMS (within the high stellar mass bin). Filled circles indicate the position of the galaxies along the tracks at a lookback time of 10.5 (empty circles) and 7.5 Gyr (filled circles). The grey line corresponds to the SFMS a $\rm z=0$ resulting from the analysis described in \ref{sec:an_SFMS}, while the dashed grey line corresponds to the SFMS analogously computed at $\rm z \sim 1$. We first note that the galaxies move in tracks that are not parallel to the SFMS at z=0. At early times, we observe that the tracks are almost horizontal, although with a slight decline. In this regime, the stellar mass builds up rapidly (see Fig. \ref{fig:cum_mass_mh} and  \ref{fig:cum_mass_sfms}) with the galaxies having high SFRs (see Fig. \ref{fig:sfh_mh} and \ref{fig:sfh_sfms}). Then, there is a sudden quench of the star formation rate in which the SFR decrease abruptly. Until this phase, the general behavior for the different bins is very similar. After these phase, however, there are tracks in which galaxies keep quenched (or maintain very low levels of SFR) until $\rm z=0$, and another ones that rejuvenate. We find that galaxies in the highest mass halos (light blue) and galaxies in the upper MS (light green) are the ones that rejuvenate the most, as we observed already in their SFHs (see Fig. \ref{fig:sfh_mh} and \ref{fig:sfh_sfms}). Galaxies in less massive halos (dark blue) and quiescent galaxies (dark green) are the ones that quench more rapidly and have lower levels of SFR at $\rm z=0$.

In this sense, for a given stellar mass, we observe that galaxies in less massive halos 
and quiescent galaxies are the ones that assemble the bulk of their stellar mass more rapidly and earlier. They also have a high SFR in the past, which declines rapidly with time (quenching). Hence, these galaxies end up with very low levels of SFR today. In contrast, galaxies in more massive halos and star-forming galaxies build the bulk of their stellar mass over more prolonged periods of time. They have lower SFRs at earlier times compared to galaxies in less massive halos and quiescent galaxies. Then, they quench, although more gradually, and rejuvenate, ending up with high SFRs today.

\section{Discussion}
\label{sec:discussion}
In this section we discuss the potential role of dark matter halo evolution in regulating galaxy stellar populations, star formation histories and the scatter across the $\rm SFR-M_{\star}$ relation, following the discussion presented in Paper I.

\subsection{Stellar-to-halo mass relation}
\label{disc:SHMR}
In section \ref{res:sfh_mh} we found that the star formation histories and cumulative mass growth of galaxies over cosmic time depend on the mass of their host dark matter halos (at fixed $M_{\star}$), with galaxies assembling earlier and faster in less massive halos, and galaxies hosted by more massive halos forming the bulk of their stars over longer periods of time, specially for the low stellar mass regime. In Paper I we also found that average stellar population properties across the stellar-to-halo mass relation (SHMR) depend both on $M_{\star}$ and $M_{h}$. At fixed $M_{\star}$, galaxies are younger and more metal-poor in more massive halos, while galaxies in less massive halos are older and more metal-rich. At the same time, we also found that, at fixed $M_{h}$, the scatter of the SHMR correlates with galaxy ages and [M/H], especially at low halo masses ($M_{h}<10^{13.2} \, M_{\odot}$), with galaxies being older and more metal-rich as $M_{\star}$ increases. In Paper I we hypothesize that these correlations are driven by the assembly time of the host dark matter halos of the galaxies. It is found both in semi-analytical models and hydro cosmological simulations that the scatter of the SHMR correlates with halo formation time (or concentration)  \citep[e.g.,][]{2013MNRAS.431..600W,2017MNRAS.465.2381M,2017MNRAS.470.3720T,2018ApJ...853...84Z,2018MNRAS.480.3978A,2019MNRAS.490.5693B,2020MNRAS.491.5747M,2021arXiv210512145C}. At fixed at fixed $M_{h}$, more massive galaxies are hosted by halos that formed earlier. Complementary, we found that at fixed halo mass, more massive galaxies are older, more metal-rich, formed faster and show lower levels of SFR at z=0. Using integral field spectroscopic data from MANGA, \citet[][]{2022ApJ...933...88O} also finds that more massive quiescent galaxies (at fixed $M_h$) are older and form over shorter time-scales (probed by the [Mg/Fe] ratio). A natural explanation for these simulated and observed trends of galaxies is therefore that galaxy ages tightly map halo formation times (with older galaxies being hosted by earlier-formed halos), which is also proposed by \citet[][]{2022ApJ...933...88O} as a possible scenario to explain their results. Under this interpretation, the scatter in the SHMR probes galaxies/halos at different evolutionary stages, and has direct implications on the observed properties of galaxies. 

On the other hand, the relation between galaxy colors or morphologies and the scatter of the SHMR is still under debate \citep[see section 6.1 of ][]{2018ARA&A..56..435W}. We would expect that the scatter of the SHMR would also correlate with galaxy colors, with more massive galaxies having redder colors, typically associated to older stellar populations, than less massive galaxies at fixed $M_h$. Our results are in agreement with \cite{2016MNRAS.455..499L}, who also studied SDSS galaxies and using halo masses from the group catalog from \citet{2007ApJ...671..153Y} and found that fixed $M_{\star}$, galaxies with higher $M_{\star}/M_{h}$ (lower $M_{h}$) tend to also have redder colors, are more quenched in star formation and are more bulge-dominated. However, using galaxy-galaxy lensing \cite{2016MNRAS.457.3200M} found that, at fixed $M_{\star}$, galaxies in less massive halos have bluer colors, which is also found for the hydro-simulation SIMBA by \citet{2021arXiv210512145C}. \cite{2020MNRAS.499.3578C} also used SDSS galaxies and halo masses from the group catalog from \citet{2007ApJ...671..153Y} and found that at fixed $M_h$ disc galaxies tend to be more massive than less massive galaxies, and these authors also find similar results for the EAGLE hydro-simulation. At first glance, these observational results seem to be in disagreement with ours, although we stress the difference in the halo mass and stellar mass measurements. Also note that we restricted our analysis to galaxies in groups with more than 3 members in order to better constrain the halo masses, although further improvement on halo mass determinations for nearby galaxies would bring light into this matter, as we discuss below. Additionally, the trends of stellar population properties across the SHMR are complex and it is hard to capture their behavior using 1D projections of it, i.e, analyzing bins of halo mass or stellar mass as it is done in these studies. For example, we showed in Fig. \ref{fig:sfh_mh} that the differences in the SFHs for galaxies in halos of different masses are more significant for the low halo mass regime, and therefore for the low stellar mass bins, consistently with the fact that we also found that the scatter of the SHMR correlates with galaxy ages and [M/H] in this regime.

\subsection{The $SFR-M_{\star}$ relation}
 The information encoded in the star-forming main sequence and its physical origin are currently a matter of debate. Some authors claim that the MS is fundamental relation in which galaxies fluctuate on short time-scales as a result of the stochasticity of star formation \citep[e.g.,][]{2010ApJ...721..193P,2013ApJ...770...57B}, implying that galaxies with similar $M_{\star}$ have also experienced similar evolutionary paths. In contrast, other authors state that this relation just reflects the average population of galaxies at a certain time \citep[e.g.,][]{2013ApJ...770...64G,2014arXiv1406.5191K,2015ApJ...801L..12A,2016ApJ...832....7A}, meaning that galaxies of the same $M_{\star}$ could have experienced very different star formation histories. In this sense, the scatter of the SFMS could be driven by systematic offsets due to different physical mechanisms, or just by stochastic effects.

In this work, we found that galaxies with different positions relative to the SFMS have also different cumulative mass growth distributions and star formation histories, specially at the low stellar mass regime (see section \ref{res:SFMS_sfh}). In this sense, at fixed stellar mass, the scatter of the SFMS seems (i.e., present day SFRs) to carry information about the evolutionary histories of the galaxies. Galaxies with low SFR today form earlier and on shorter time-scales that galaxies with high SFRs today, which form the bulk of their stars over longer periods of time,  qualitatively in agreement with results predicted by the EAGLE simulation \citep[e.g.,][]{2019MNRAS.484..915M}. 

We discussed above that potential role of halo evolution driving the scatter of the SHMR, with different theoretical works finding that more massive galaxies are hosted by earlier formed halos than less massive galaxies at fixed halo mass (see section \ref{disc:SHMR}). At the same time, \citet[][]{2019MNRAS.484..915M} also found that the scatter of the $\rm SFR-M_{\star}$ relation correlates with halo formation time for simulated galaxies from EAGLE (at fixed stellar mass and for low stellar masses), with high SFRs galaxies being hosted by later formed halos than galaxies with lower SFRs, which are hosted by earlier formed halos. In this regard, \citet[][]{2019MNRAS.484..915M} already proposed that the scatter of the SHMR might be connected to the scatter of the $\rm SFR-M_{\star}$ relation. Hence, taking into account these findings and our proposed scenario in which halo formation time potentially drives stellar populations and SFHs across the SHMR, we speculate that galaxies at different positions relative to the SFMS are also at different evolutionary stages. In this picture, galaxies with high SFRs today, which have more extended SFHs, are hosted by later-formed halos, being at a later evolutionary stage than galaxies with lower SFRs at present day, which assembled in earlier-formed halos, and on shorter time-scales and at earlier times. And as mentioned before, these earlier-formed galaxies have older and more metal-rich stellar populations at present day than later-formed galaxies. \\

\subsubsection{Scatter of the SFMS at different times}
Our fossil records approach can also provide information about the scatter of the SFMS at different times, given that, by definition, a slice of a galaxy SFH at a given time indicates the SFR of the galaxy at that time.

  Looking at the different SFHs shown in this work (Fig. \ref{fig:sfh_mh} and \ref{fig:sfh_sfms}) for a given stellar mass, we observe that galaxies with high SFRs today had lower SFR at earlier times, while galaxies with low SFRs today tend to have higher SFRs at early epochs. At the same time, we have seen that upper MS galaxies (high SFRs today) have experienced a different evolutionary histories than quiescent galaxies (low SFRs today), with the latter galaxies forming the bulk of their stars earlier. Hence, this could imply that the nature of the scatter of the SFMS (i.e., its evolutionary component) seems to change with time, as star-forming galaxies today (high present-day SFRs) populate the upper part of the MS, while they had lower SFRs in the past than quiescent galaxies. Moreover, we also observe that the different SFHs cross at a given time, suggesting that the scatter of the $\rm SFR-M_{\star}$ relation would be minimum at the time in which they cross. At earlier times and after this crossing point, these SFHs also suggest that there is an `inversion' of the scatter of the SFMS with respect to z=0 in the sense that quiescent galaxies today are the ones with the higher SFRs and MS galaxies today are the ones with the lower SFRs. However, we note that we are only fixing the stellar mass of the galaxies at $z=0$, and at a certain LBT the SFHs does not necessarily correspond to galaxies of same mass at that time. Hence, in order to bring light into this matter it would be needed to study the evolution of the SFMS with time by studying the scatter of the SFMS of galaxies at different epochs.

\subsubsection{Galaxy Quenching}
It is often inferred that star-forming galaxies at higher-z are the progenitors of the quiescent galaxies that we observe in the Local Universe, and thus, have experienced a quenching process that ceases their star formation and have kept them quiescent with cosmic time (see section \ref{sec:intro}). In contrast, our fossil record approach allows us to trace individual galaxy evolutionary histories, and hence, this quenching process without further assumptions on the galaxy populations at different times. 

We clearly observe this quenching process in the galaxy SFHs as a rapid decline of the SFR at early times, specially for massive galaxies (Fig. \ref{fig:sfh_mh} and \ref{fig:sfh_sfms}). Additionally, the SFHs of quiescent galaxies also show this quenching phase after which they keep low levels of SFR, with these SFHs resembling the ones of massive galaxies of Fig. \ref{fig:sfh_mh}, and the ones of galaxies in less massive halos (for the low stellar mass regime). While in the case of star-forming galaxies, they rejuvenate after a more gradual decline of their SFR. 

This quenching phase is more obvious in the tracks of quiescent galaxies across the $\rm SFR-M_{\star}$ relation (Fig. \ref{fig:quenching}), which show that these galaxies grow in mass until a certain time, after which their SFRs decrease substantially with little or negligible mass growth during the process (as the tracks are almost vertical). In addition, galaxies in the most massive stellar bins, although not shown in this figure, also have a similar behavior. These tracks are in agreement with the ones of quiescent galaxies shown in \citet{2019MNRAS.487.4939M,2022MNRAS.513L..10M,2022ApJ...926..134T}. 

Following the previous discussion about halo evolution, we propose that galaxies at earlier formed-halos, are the ones that are more massive, have old and metal-rich stellar populations, quench earlier, as they also form early on and on short time-scales, with their SFHs showing this quenching phase. In addition, it is also predicted by hydro-simulations that halo assembly history influences the star formation histories of the galaxies \citep{2021MNRAS.501..236D}. They found that earlier-formed halos host quenched galaxies and more massive central black holes, which was also predicted by previous theoretical studies \citep{2010MNRAS.405L...1B,2011MNRAS.413.1158B}, and hence, inject more energy by AGN feedback, effectively quenching the galaxies. Therefore, in this scenario the quenching of massive galaxies would be driven as a combination of halo assembly and black hole growth.

\subsection{Caveats and limitations}
\label{sec:disc_cav}
\subsubsection{Time resolution of spectral synthesis models and the `archaeological' approach}
The fossil record approach is a powerful that allows to derive the individual SFHs of the galaxies (see section \ref{sec:intro}). Nevertheless, we also have to take into account that this method has caveats and systematic effects associated to it in order to avoid the overinterpretation of our findings. 

First, we note that SFHs derived with full-spectral fitting can only be recovered up to a certain level on reliability, as biases on the SSP models extend to the resulting SFHs. In particular, one of the main caveats of the method is that the time resolution of the SFHs is determined by stellar evolution. Specially at older ages, the age resolution decreases due the degeneracy in the features of SSP models with similar ages, given that changes in the spectra of young stars occur in shorter timescales than in the case of old stellar populations. Therefore, the time resolution of the SSP models determines the star formation events that can be measured in the SFHs, with our method being more sensitive to the ones of young stellar populations. In this sense, the star formation bursts that we observe at later times in our SFHs are probably also present at earlier times, although we do not have enough resolution to distinguish them. Moreover, this also implies that our results are also sensitive to the stellar libraries and SSP models in the fitting procedure. For example, SSP models of very young stellar populations are less reliable given that the modelling of these very young stars is more complex, hence, making the SFRs associated to these SSPs less precise.

Despite the aforementioned limitations of the models, we note that the fossil record approach is able to recover reasonably well non-parametric galaxy SFHs, and it has been thoroughly tested and applied to a variety of cases. In particular, \citet[][]{2015A&A...583A..60R,2018A&A...617A..18R} found that SFHs derived with STECKMAP are in good agreement with those measured using color-magnitude diagrams. More recently, \citet[][]{2020ApJ...896...13B} showed how pPXF can be used to recover the SFH and chemical time-evolution of galaxies based on integrated spectra, which are also consistent with the corresponding analysis of resolved stars (their Fig. 9). Additionally, \citet[][]{2020MNRAS.491..823B} also found that pPXF can recover the mass distribution of stellar populations in the age-metallicity space from integrated mock spectra using the EAGLE hydro-simulation \citep[][]{2015MNRAS.446..521S}. Moreover, although this method does not provide precisely in time the corresponding bursts of star formation of the SFHs due to the limited SSP ages, it is able to robustly obtain the overall shape of the SFHs. In fact, \citet[][]{2019MNRAS.483.4525I} are able to recover reasonably well the SFHs of the mock MaNGA galaxies using PIPE3D \citep[][]{2016RMxAA..52...21S}, although the SFHs are significantly smoothed at old ages due to the age sampling of the SSPs. More recently, \citet{2022MNRAS.515..320N} are able to recover as well the overall shape of the SFHs of mock MaNGA galaxies using FIREFLY \citep[][]{2017MNRAS.472.4297W}. Similar is the case of spectral energy distribution (SED) fitting techniques, which are also able to accurately recover the overall shape of the SFHs, being more difficult to robustly derive  higher-order quantities such as burst  fraction or quenching time \citep[][]{2019ApJ...876....3L,2022ApJ...935..146S}. 

 Additionally, it has to be taken into account that full-spectral fitting codes  perform worse with increasing noise in the spectra, and therefore, a lower S/N translates into more scatter and biases in the recovered parameters \citep[][]{2006MNRAS.365...46O,2007MNRAS.381.1252T}, as the noise washes away differences associated to different stellar templates. Additionally, the spectra of young stellar populations introduce larger biases and scatter than the ones of older stellar populations \citep[][]{2007MNRAS.381.1252T,2018MNRAS.478.2633G}. In particular, when young stellar populations dominate the galaxy spectra at low S/N, the mass associated to old stellar populations derived with pPXF can be underestimated \citep[][]{2018MNRAS.478.2633G}.  However, we have checked that in our sample the flux associated to the youngest stellar populations (ages < 1 Gyr) at lowest stellar mass bin (where this effect is expected to be higher, as seen in Fig. 6) is approximately 10\% of the or less of the total optical flux, as measured by the best-fitting pPXF weights. Therefore, even for low stellar masses, we do not expect masses to be unrealistically underestimated.

On the other hand, this method can not distinguish between stellar populations formed in-situ or ex-situ, and hence, we can not determine the mass fraction of a galaxy originated from mergers or accreted systems. Thus, the individual SFHs of a galaxy that we derive with this method can mix the stellar content of several galaxies, and we can misinterpret it as the evolution of a single galaxy. However, we note that we alleviate this concern by studying the average behavior of galaxies with similar properties (e.g., stellar mass or halo mass) instead of galaxies on an individual basis. Moreover, the SDSS fiber (3'') probes the inner few kpc of the galaxies (an average of 5 kpc for our galaxy sample). These inner regions are expected to be dominated by in-situ formed material, as revealed by the tight scaling relations between stellar population properties and stellar mass. 

Finally, we note that throughout this paper series we do not aim to set absolute constraints on the t50 values or in any other stellar population parameter (see Paper I). Instead, we are comparing the stellar population properties and SFHs of galaxies in a relative way, as we are interested in the relative difference of these observables for different galaxies. This relative approach is much more robust, in particular when paired with empirical stellar population synthesis models, and has proven to be systematically sensitive to sub-Gyr age variations for even old stellar populations \citep[see e.g. Fig.7 in  ][]{2018MNRAS.475.3700M}. Moreover, we also note that in this work we have employed a relative narrow spectral range (4800–5400 \AA) to derive the SFHs, which we selected for consistency with Paper I. Nonetheless, we have tested the reliability of results by repeating the whole analysis shown in this paper extending the wavelength range to 3800-5800 \AA, and we found that coherent results and in agreement with the ones shown in sections \ref{res:sfh_mh} and \ref{res:SFMS_sfh}. Thus, we demonstrated that this wavelength range also allow us to confidently constrain ages and SFHs, as expected, given that age variations leave a distinct and measurable imprint across the whole optical range \citep[see e.g. Fig. 6 in ][]{2006MNRAS.365...46O}.

\subsubsection{Halo mass estimations}
As we already discussed in Paper I, one important limitation of our analysis is the halo mass determination. At the moment, there are not direct halo masses measurements (e.g., $M_h$ estimated through galaxy rotation curves) available for large samples of galaxies. Therefore, in order to assess the role of halo mass regulating the evolution of nearby galaxies, we used indirect halo mass estimations drawn from the group catalog described in \citet{2007ApJ...671..153Y}, which use abundance matching to populate observed galaxies into halos from dark matter only simulations by ranking the total luminosity (or mass) of the groups. Although these halo masses are obtained indirectly, we have checked that we recover the general trends of the stellar population properties across the SHMR and VDHMR using other group catalogs (see Paper I). 

At the same time, there are other approaches to estimate halo masses, such as satellite kinematics or weak-lensing (although they also have associated their own uncertainties and systematic effects), and quantities that scale with halo mass, such as the number of globular cluster (GC) systems.

\section{Conclusions}
\label{sec:concl}
We have investigated the time evolution of the stellar population properties of nearby galaxies. In particular, we derived the cumulative mass fraction distributions and the star formation histories of these galaxies and studied them in terms of their host dark matter halos and of the scatter of the star-forming main sequence. Our results can be summarized as follows:

(i) We found that the cumulative mass fraction distributions and star formation histories of the galaxies depend on the mass of their host halos. At fixed stellar mass, galaxies form the bulk of their stars earlier and faster as halo mass decreases, specially for low stellar masses. 

(ii) Galaxies have different star formation histories depending on their position relative to the SFMS at z=0, with high SFR galaxies at present day forming over longer periods of time than the ones with lower SFRs, which form earlier and on shorter time-scales. 

(iii) We interpret that halo formation time drives galaxy star formation histories and the scatter across the $\rm SFR-M_{\star}$ relation, with earlier-formed halos hosting galaxies with low SFRs today and have assembled at early epochs and on short time-scales. 

Our results indicate that halo mass modulates the baryonic cycle of galaxies, and in particular, galaxy quenching. Although galaxy mass or velocity dispersion are found to be the main drivers of the stellar population properties, halo assembly and black hole growth might have secondary, but key roles regulating the stellar content and star formation histories of galaxies. To confirm this scenario, it would be necessary to better constrain the halo masses of the galaxies, either with direct halo mass measurements, or with another halo mass estimations or proxys (e.g., lensing, number of GCs systems). In addition, observational estimations of other halo properties such as halo formation time for galaxies in the Local Universe would be key to further probe the role of dark matter halos in regulating their evolution. On the other hand, spectroscopic observations of higher redshift galaxies will provide galaxy SFHs with a better time resolution a better sensitivity to the oldest stellar populations present in the galaxies.

\section*{Acknowledgements}
 We acknowledge support through the RAVET project by the grant PID2019-107427GB-C32 from the Spanish Ministry of Science, Innovation and Universities (MCIU), and through the IAC project TRACES which is partially supported through the state budget and the regional budget of the Consejer\'ia de Econom\'ia, Industria, Comercio y Conocimiento of the Canary Islands Autonomous Community. IMN also acknowledges support from grant ProID2021010080 in the framework of Proyectos de I+D por organismos de investigaci\'on y empresas en las \'areas prioritarias de la estrategia de especializaci\'on inteligente de Canarias (RIS-3). FEDER Canarias 2014-2020.

We are grateful to Marc Huertas-Company for his advice and comments. We thankfully acknowledge the technical expertise and assistance provided by the 
  Spanish Supercomputing Network (Red Española de Supercomputación) and the Instituto de Astrofísica de Canarias (IAC), as well as the computer 
  resources used: the LaPalma Supercomputer, and the High Performance Computers Diva and Deimos, located at the IAC.

This research made use of Astropy, a community-developed core Python package for Astronomy \citep[][]{astropy:2013,astropy:2018}, and of the Numpy \citep[][]{2020Natur.585..357H}, Scipy \citep[][]{2020SciPy-NMeth} and Matplotlib \citep[][]{2007CSE.....9...90H} libraries.

Funding for SDSS-III has been provided by the Alfred P. Sloan Foundation, the Participating Institutions, the National Science Foundation, and the U.S. Department of Energy Office of Science. The SDSS-III web site is http://www.sdss3.org/.

SDSS-III is managed by the Astrophysical Research Consortium for the Participating Institutions of the SDSS-III Collaboration including the University of Arizona, the Brazilian Participation Group, Brookhaven National Laboratory, Carnegie Mellon University, University of Florida, the French Participation Group, the German Participation Group, Harvard University, the Instituto de Astrofisica de Canarias, the Michigan State/Notre Dame/JINA Participation Group, Johns Hopkins University, Lawrence Berkeley National Laboratory, Max Planck Institute for Astrophysics, Max Planck Institute for Extraterrestrial Physics, New Mexico State University, New York University, Ohio State University, Pennsylvania State University, University of Portsmouth, Princeton University, the Spanish Participation Group, University of Tokyo, University of Utah, Vanderbilt University, University of Virginia, University of Washington, and Yale University.

\section*{Data Availability}
The SDSS group and cluster catalog from \citet{2007ApJ...671..153Y} is available at https://gax.sjtu.edu.cn/data/Group.html. The SFR catalog from \citet[][]{2004MNRAS.351.1151B} is available at https://wwwmpa.mpa-garching.mpg.de/SDSS/DR4/. The stellar population properties and SFHs of the galaxies are available on reasonable request to the corresponding author. 

\bibliographystyle{mnras}
\bibliography{references} 

\appendix
\section{Selection of star-forming galaxies to derive the SFMS}
\label{ap:sel_sf}

\begin{figure*}
    \centering
    \includegraphics[scale = 0.48]{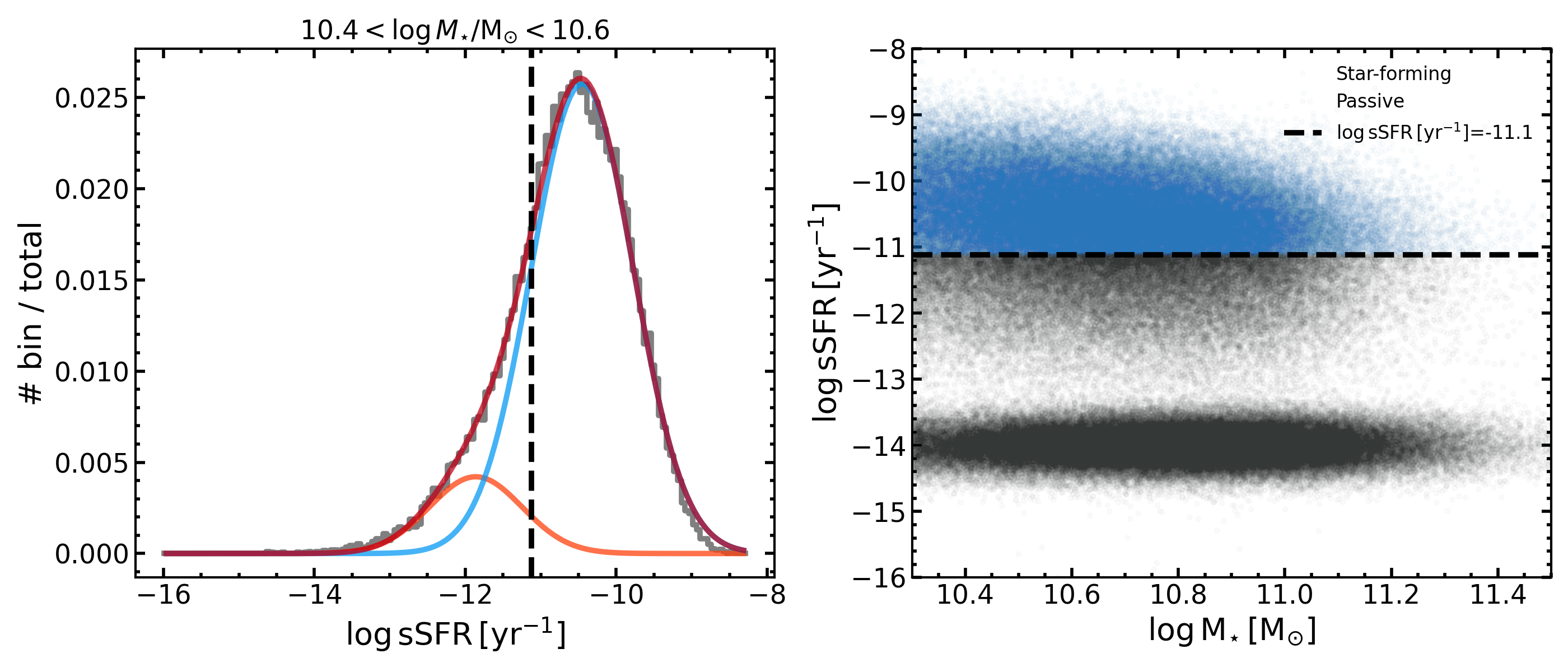}
    \caption{\textit{Left:}  Specific star formation rate distribution of the galaxies in a narrow stellar mass bin ($10^{10.4}$<$M_{\star}$<$10^{10.6}$). The grey solid line shows the normalized sSFR distribution, while the the blue and orange lines show the two gaussian components of the best-fitting model, which is shown as a red solid line. The vertical black dashed line indicates our sSFR threshold to select star-forming galaxies. \textit{Right}: Specific star formation rate as a function of stellar mass. 
    The sSFR threshold of $\rm \log sSFR = -11.1$ used to select star-forming galaxies is shown as a black dashed line. Individual galaxies are shown with circles, with star forming galaxies shown in blue, and the rest of the galaxy population shown in dark grey. For visualization purposes, quiescent galaxies with $\rm SFR_{70 \ Myr}=0$ are artificially plotted with arbitrarily  low sSFR  values of $\rm \log sSFR = 14$ dex randomly scattered with a standard deviation of 0.25 dex.}
    \label{fig:def_sf}
\end{figure*}

In order derive the SFMS, we first need to select star-forming galaxies. For that, we select galaxies with SFR > 0, but also impose a sSFR threshold cut with the aim to avoid the tail of quenched galaxies. To select this sSFR threshold, we first choose the galaxies in a narrow stellar mass bin centered in $\rm \sim10^{5} \,  M_{\odot}$. Then, we fit the sSFR distribution of these galaxies with a model comprised by with two gaussian components. The left panel of Fig. \ref{fig:def_sf} shows the normalized sSFR  distribution with a grey line, the blue and orange lines show the two gaussian components of the best-fitting model, which is shown as a red solid line. The blue line corresponds to the distribution of star-forming galaxies, as it is centered at higher sSFRs, while the one of passive galaxies is shifted towards less sSFR. Then, we select our sSFR threshold as the mean of the gaussian component corresponding to the star-forming population (blue line) less 1$\rm \sigma$, which is shown as a vertical black dashed line in the left panel of Fig.\ref{fig:def_sf}. Thus, we define a galaxy as star-forming when it has an sSFR above $\rm 10^{-11.1} \, yr^{-1}$. We observe that the number of passive galaxies shown in this figure is considerably lower than the one of star-forming galaxies. This is because our method is not sensitive to very low levels of SFR, which effectively translate into no recent star formation ($\rm SFR_{70\, Myr}$=0), and hence passive galaxies do not appear in this plot by construction.  For visual purposes we artificially plot these galaxies in the right panel of Fig. \ref{fig:def_sf}. Given the low number of passive galaxies with SFR > 0, we also repeated the analysis but selecting as star-forming galaxies all the ones with SFR > 0 (without a imposing a sSFR cut). With this test, we find that our results are not very sensitive to tail of quenched galaxies, as we recover the results presented in section \ref{res:SFMS_sfh}. 

We apply this sSFR threshold to the whole sample to select star-forming galaxies. Note that this selection of star-forming galaxies is only used for the SFMS determination. In right panel of Fig. \ref{fig:def_sf} we show the star-forming galaxies and passive galaxies across the sSFR-$\rm M_{\star}$ relation. Star-forming and passive galaxies are shown with blue and grey circles, respectively. The sSFR threshold above we classify galaxies as star-forming is shown as an horizontal black dashed line. By construction, galaxies without recent star-formation ($\rm SFR_{70\, Myr}$=0) do not appear in this figure. Hence, for visual purposes, we artificially plot these galaxies scattered randomly around a center arbitrarly low sSFR of $\rm 10^{-14} \, yr^{-1}$ with a standard deviation of 0.25 dex. Note that we do not use these values for any computations, but just for visualization.

\bsp	
\label{lastpage}
\end{document}